\documentclass[sigconf]{acmart}
\AtBeginDocument{%
  }

\setcopyright{acmlicensed}
\copyrightyear{2018}
\acmYear{2018}
\acmDOI{XXXXXXX.XXXXXXX}
\acmConference[Conference acronym 'XX]{Make sure to enter the correct
  conference title from your rights confirmation email}{June 03--05,
  2018}{Woodstock, NY}
\acmISBN{978-1-4503-XXXX-X/2018/06}

\usepackage{graphicx}   
\usepackage{subcaption} 
\usepackage{svg} 

\usepackage{enumitem}
\usepackage[normalem]{ulem}
\usepackage{soul}
\usepackage{booktabs}
\usepackage{multirow}
\usepackage{longtable}
\usepackage{pgfplots}
\pgfplotsset{compat=1.18}

\usepackage{makecell}
\usepackage{tikz}
\usepackage{booktabs}
\usepackage{array}
\usepackage{longtable}
\usepackage{geometry}
\usepackage{soul}
\usepackage{multicol}

\definecolor{DMColor}{HTML}{fa8072}
\definecolor{BColor}{HTML}{91908f}
\definecolor{purplehl}{RGB}{142, 124, 195}
\definecolor{cyanhl}{RGB}{118, 165, 175}
\definecolor{orangehl}{RGB}{255, 153, 0}
\definecolor{pinkhl}{RGB}{194, 123, 160}

\newcommand{\toolname}{\textsc{DesignMentor}}

\newcommand{\pnk}[1]{\textcolor{pinkhl}{#1}}
\newcommand{\org}[1]{\textcolor{orangehl}{#1}}
\newcommand{\cyn}[1]{\textcolor{cyanhl}{#1}}
\newcommand{\pur}[1]{\textcolor{purplehl}{#1}}

\begin{document}

\title{From Answer Givers to Design Mentors: Guiding LLMs with the Cognitive Apprenticeship Model}

\author{Yongsu Ahn}
\affiliation{%
  \institution{Boston College}
  \city{Chestnut Hill}
  \state{Massachusetts}
  \country{USA}}
\email{anyon@bc.edu}

\author{Lejun R Liao}
\affiliation{%
  \institution{Boston College}
  \city{Chestnut Hill}
  \state{Massachusetts}
  \country{USA}}
\email{liaolc@bc.edu}

\author{Benjamin Bach}
\affiliation{%
  \institution{Inria}
  \city{Bordeaux}
  \country{France}
}
\affiliation{%
  \institution{University of Edinburgh}
  \city{Edinburgh}
  \country{United Kingdom}
}
\email{benjamin.bach@inria.fr}

\author{Nam Wook Kim}
\affiliation{%
  \institution{Boston College}
  \city{Chestnut Hill}
  \state{Massachusetts}
  \country{USA}}
\email{nam.wook.kim@bc.edu}

\renewcommand{\shortauthors}{Trovato et al.}

\begin{abstract}
 Design feedback helps practitioners improve their artifacts while also fostering reflection and design reasoning. Large Language Models (LLMs) such as ChatGPT can support design work, but often provide generic, one-off suggestions that limit reflective engagement. We investigate how to guide LLMs to act as design mentors by applying the Cognitive Apprenticeship Model, which emphasizes demonstrating reasoning through six methods: modeling, coaching, scaffolding, articulation, reflection, and exploration. We operationalize these instructional methods through structured prompting and evaluate them in a within-subjects study with data visualization practitioners. Participants interacted with both a baseline LLM and an instructional LLM designed with cognitive apprenticeship prompts. Surveys, interviews, and conversational log analyses compared experiences across conditions. Our findings show that cognitively informed prompts elicit deeper design reasoning and more reflective feedback exchanges, though the baseline is sometimes preferred depending on task types or experience levels. We distill design considerations for AI-assisted feedback systems that foster reflective practice.

\end{abstract}

\ccsdesc[500]{Human-centered computing~Empirical studies in interaction design}
\ccsdesc[500]{Human-centered computing~Empirical studies in visualization}
\ccsdesc[500]{Human-centered computing~Natural language interfaces}

\keywords{Design Feedback, Design Reasoning, Reflective Practice, Metacognition, AI-Assisted Design}
\begin{teaserfigure}
  \centering\includegraphics[width=.9\textwidth]{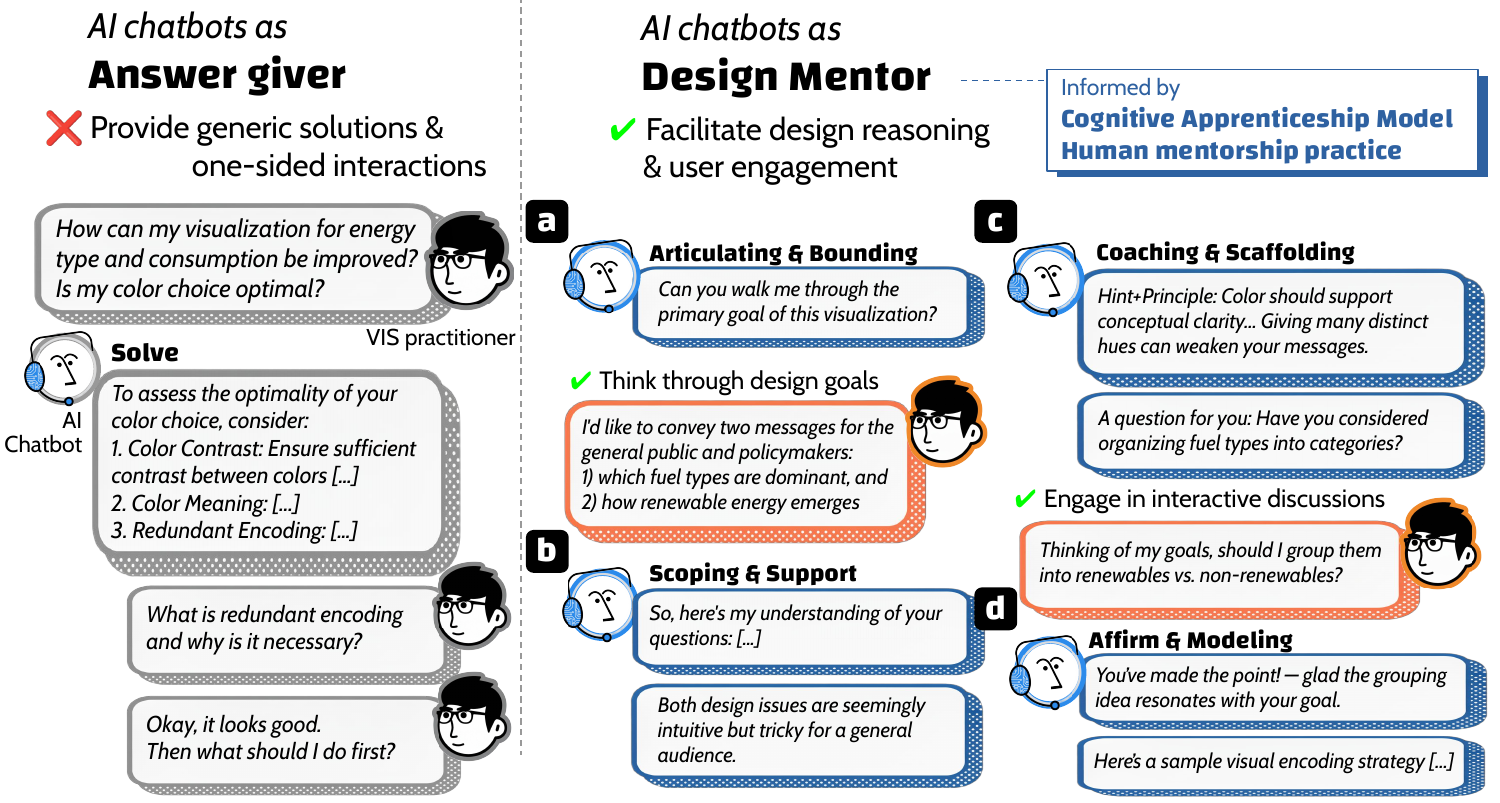}
  \vspace{-1em}
  \caption{Overview of \toolname{}'s approach to transforming AI chatbots from answer givers to design mentors. In the conversations with AI chatbots, a visualization practitioner seeks to improve the m visualization. As a conventional answer giver, a typical AI chatbot responds with a generic checklist, leaving the user to ask follow-up questions without deliberate support. Informed by the Cognitive Apprenticeship Model and human mentorship practices, \toolname{} guides the practitioner through a structured feedback process: (a) In Articulating \& Bounding, the mentor prompts the user to verbalize their design goals and intent. (b) Through Scoping \& Support, the mentor restates the user's questions while acknowledging the challenge. (c) In Coaching \& Scaffolding, the mentor diagnoses the current design and offers principle-based hints—leading the user to their own insight about grouping renewables vs. non-renewables. (d) Finally, through Affirm \& Modeling, the mentor validates the user's reasoning before offering a sample encoding strategy to foster reflection and active design reasoning.\color{black}}
  \Description{Enjoying the baseball game from the third-base
  seats. Ichiro Suzuki preparing to bat.}
  \label{fig:teaser}
\end{teaserfigure}


\maketitle

\section{Introduction}
Design feedback serves as a cornerstone of creative practice, enabling practitioners to refine their work through external perspective and expert guidance~\cite{alabood2023systematic,gray2013informal}. While it helps them identify flaws and suggest improvements of their design artifacts, its deeper value lies in catalyzing designer's own cognitive development and design reasoning~\cite{yen2017listen,mccall2012critical,hokanson2012design}. At its core, effective design feedback facilitates what Schön described as ``reflection-in-action''~\cite{schon2017reflective}, where practitioners reflect on their own design artifacts, explore potential solution spaces, and refine their approach based on targeted critique. 

Recent advances in Large Language Models (LLMs) have shown both promises and perils of providing feedback on design processes and artifacts for practitioners. As previous studies revealed~\cite{kim2023good,ahn2025understanding,duan2024generating,shin2025visualizationary}, ChatGPT can provide broad and generally useful suggestions, acting as a supportive design assistant that helps practitioners address design problems. However, they tend to present generic, one-size-fits-all suggestions as a response to a given query~\cite{kim2023good,duan2024generating}. Despite their extensive knowledge and fluency, people often perceive LLMs as transactional solution generators, limiting interactions to a simple question-and-answer exchange~\cite{kim2024chatgpt,kobiella2024if}. This one-sided interaction can further hamper practitioners’ own reflective thinking on their design decisions and reasoning behind the decisions. 

On the other hand, research has suggested that feedback processes in general go beyond simply offering critiques and suggestions, arguing for deliberate feedback mechanisms that prompt one’s reflection, make their thinking visible, and help them iteratively build and reorganize their understanding~\cite{roldan2021pedagogical,cook2020designing,adams2017approaches}. As decades of educational research have demonstrated, when carefully designed and applied, such feedback can facilitate deeper cognitive engagement across diverse domains and tasks~\cite{hattie2007power}. While such mentorship is not readily available due to lack of human resources and training~\cite{jorke2025gptcoach,arakawa2024coaching}, tools designed to externalize and scaffold individuals' thought process can enable structured and guided feedback, allowing designers to articulate their reasoning, revisit earlier assumptions, and identify conceptual gaps~\cite{huang2025ai,zhang2020reflectionscope}.

In this work, we investigate how we can guide LLMs to move beyond simply presenting design solutions to serve as design mentors, enabling practitioners to become active participants in their own reflective design processes rather than passive recipients of advice. To address this challenge, we draw on the Cognitive Apprenticeship Model (CAM)~\cite{collins2018cognitive,brown1989situated}, which adapts traditional apprenticeship---where learners work alongside skilled practitioners---to teach cognitive tasks using six instructional methods such as modeling, coaching, scaffolding, articulation, reflection, and exploration. Our formative study finds that AI chatbots' limited capability as merely advice-givers was perceived as a significant challenge practitioners face \ref{fig:teaser}, such as generic and ready-made solutions and one-sided interactions. We discover opportunities from the analysis that human mentors actively engage in the comprehensive feedback practices suggested in CAM and even beyond, which can serve as design mentorship guidelines to instruct AI chatbots in providing more guided feedback process and fostering design reasoning. 


Based on our formative findings, we propose \toolname{}, a cognitively-informed AI feedback system that operationalizes the principles of cognitive apprenticeship through structured prompting (Figure \ref{fig:teaser}. Our system builds on a design mentorship codebook derived from analyzing human expert behaviors, extending CAM's framework with additional mentorship strategies contextualized to design feedback processes. Through an iterative prompt refinement involving pilot testing with visualization practitioners, we developed seven design guidelines that guide \toolname{} to facilitate structured, multi-phase feedback sessions using a custom GPT implementation. To test this, we conduct a within-subjects experiment with a diverse pool of students, data practitioners, and researchers practicing data visualization. Each participant brought their own visualization and interacted with both the baseline and the instructional LLM in randomized order, asking design-related questions for each condition. We collected follow-up surveys and interviews to capture participants’ assessments of their experiences, and analyzed conversational logs to identify the patterns of interactions, and feedback exchanges and contents. 

Our findings demonstrate that \toolname{} significantly outperformed the baseline in providing complete feedback loops, fostering metacognitive awareness, and delivering contextually relevant guidance. Participants found the structured approach more engaging and conducive to design reasoning, though it required greater cognitive effort. Importantly, we discovered that effectiveness varied by design phase---participants in exploration and development phases strongly preferred \toolname{}, while those in evaluation phases sometimes favored the baseline's directness. The contribution of our study includes:

\begin{itemize}
    \item \textbf{Design mentorship codebook and guidelines for prompt instruction.} Based on findings from our formative study, we provide a comprehensive analysis of how human mentors differ from AI chatbots in design feedback, revealing gaps in interactive, reasoning-driven support that current LLMs fail to address.
    \item \textbf{\toolname{}, a cognitively-informed AI design mentor.} We propose \toolname{}, an AI chatbot guided by seven design guidelines that operationalize the CAM through structured prompting strategies, transforming LLMs from answer-givers into process-oriented design mentors.
    \item \textbf{Empirical findings for a nuanced implementation of AI design mentor from a suite of evaluations.} Our user study with 24 visualization practitioners reveals that the effectiveness of structured AI mentorship is highly contextual, with clear preferences emerging based on design phase, user experience level, and task goals. We provide empirical evidence for when guided approaches excel versus when direct feedback is preferred, offering crucial insights for designing adaptive AI mentorship systems, with quantitative investigation on how \toolname{} transforms the feedback quality and interaction between AI and practitioners.
\end{itemize}

\section{Related Work}
\subsection{Cognitive Apprenticeship Model}
\paragraph{Background}
The CAM~\cite{collins2018cognitive} is a pedagogical framework that explains learning as guided practice, where experts support novices in developing skills and knowledge. Originally motivated from traditional apprenticeship where learners observe and participate in authentic work practices, CAM extends this approach to domains that involve cognitive abilities such as reading, writing, and problem-solving. One of its central principles is to make hidden thinking processes explicit and teachable.


\textbf{Six Feedback Methods.} The framework encompasses six feedback methods that support students in developing integrated skills through observation, guided practice, and increasing autonomy.

\begin{itemize}
    \item \textbf{Mentor-driven methods} focus on making experts' feedback and thinking more visible and gradually exposed to support mentees' tasks and skills.  \textit{Coaching} consists of observing students on their current performance and ``offering hints, feedback, reminders, and new tasks aimed at bringing their performance closer to expert performance.'' \textit{Scaffolding} provides support to help students carry out tasks, whether through suggestions, physical tools, or the teacher ``executing parts of the task that the student cannot yet manage,'' with gradual fading as competence develops. \textit{Modeling} involves experts performing tasks so students can observe and build conceptual models of required processes, requiring ``externalization of usually internal processes and activities.''
    
    \item \textbf{Mentee-driven methods} center on promoting mentees' metacognitive awareness and autonomous problem-solving capabilities. \textit{Articulation} involves methods for getting students to ``articulate their knowledge, reasoning, or problem-solving processes'' through inquiry teaching, think-aloud protocols, or peer critique roles. \textit{Reflection} enables students to ``compare their own problem-solving processes with those of an expert, another student, and ultimately, an internal cognitive model of expertise'' through replaying and analyzing performance. \textit{Exploration} pushes students toward independent problem-solving by ``setting general goals for students and then encouraging them to focus on particular subgoals of interest to them,'' representing the natural culmination of fading support. In practice, mentors play an essential role in cultivating these exploratory behaviors, as learners do not often naturally engage in such independent problem-solving without deliberate facilitation.
\end{itemize}

\textbf{Core Feedback Principles.} Three fundamental differences distinguish cognitive apprenticeship from traditional apprenticeship, forming its core principles. First, \textit{making thinking visible (Verbalize)} addresses the key challenge that ``in cognitive apprenticeship, one needs to deliberately bring the thinking to the surface, to make it visible'' since cognitive processes are naturally hidden unlike observable physical tasks. Second, \textit{situated learning (Exemplify)} involves contextualizing abstract academic tasks ``in contexts that make sense to students'' rather than remaining ``divorced from what students and most adults do in their lives.'' Third, \textit{transfer and generalization (Generalize)} requires presenting ``a range of tasks, varying from systematic to diverse'' and encouraging students to ``reflect on and articulate the elements that are common across tasks'' to develop flexible, transferable expertise. \\

The effectiveness of CAM has been demonstrated across multiple domains. It has been shown to enhance student autonomy in clinical education through reflective guidance~\cite{affendy2021impact}, support inquiry-driven learning to strengthen knowledge transfer and improve self-efficacy in STEM~\cite{koretsky2009engaging}, and foster significant gains in argumentative writing via scaffolding and self-regulation~\cite{tsiriotakis2021effects}. The model has also proven adaptable to online and blended environments, strengthening reflective and collaborative skill development~\cite{madsen2019planning, fennell2019computational}. In design practice, it has demonstrated the potential to address the core challenge of making tacit knowledge—such as heuristics and critique—accessible to learners through structured cycles of iterative feedback and reflection~\cite{geyser2024implicit, nicotra2022innovative}.

Recent advances in artificial intelligence extend the model's mentorship potential, particularly through LLMs that can provide timely, context-sensitive feedback during design sessions~\cite{yadav2025idea}. Such systems can simulate expert mentoring, fostering the bidirectional reasoning and metacognitive reflection essential for creative and technical practice. Yet these opportunities also raise questions of responsible integration: AI should augment, not replace, human judgment, requiring careful pedagogical design~\cite{huber2024leveraging}. Consequently, research increasingly explores AI-augmented feedback and networked learning environments to expand expert guidance~\cite{oztok2018socialization}, while ongoing work investigates frameworks for embedding AI within cognitive apprenticeship to sustain reflective learning and mentorship~\cite{dickey2007barriers, oktaviyanthi2023cognitive}.

Our work contributes to this line of research by operationalizing the six core methods of CAM into a concrete set of prompt instructions. We pay attention to visualization design contexts to guide an LLM to act as a design mentor, showing how our AI chatbots transform the way it facilitates design reasoning and interactions with practitioners.

\subsection{Design feedback in HCI and visualization}
Design feedback is a critical mechanism through which designers refine their work, validate decisions, and ensure alignment with user needs and design principles. Peer critique is a common way to obtain feedback, yet it can be challenging: the fear of criticism and lack of anonymity often make individuals uncomfortable receiving input from colleagues~\cite{diehl1987productivity,hui2015using}. For self-employed designers, finding peers who can provide valuable feedback may be even more difficult~\cite{xu2012you}. To address these challenges, prior work has turned to online crowdsourcing, which offers access to large and diverse participant pools, faster turnaround, and relatively low costs~\cite{buhrmester2016amazon}. Early systems like CrowdCrit demonstrated that, with structured workflows, feedback aggregated from non-expert crowd workers could approximate the quality of expert critiques~\cite{luther2015structuring}.

Yet, raw crowdsourced feedback is often limited by superficiality, sentiment bias, and a mismatch between the holistic nature of design and the fragmented microtask structure of crowd platforms. To overcome these limitations, researchers have explored scaffolding strategies: structuring tasks into micro-units~\cite{xu2014voyant,luther2015structuring}, supplying predefined rubrics~\cite{yuan2016almost}, providing demonstrative examples~\cite{kang2018paragon}, and offering explicit feedback guidelines~\cite{krause2017critique}. For instance,~\cite{lekschas2021ask} shows that prompting workers with open-ended questions, rather than declarative requests, reduces sentiment bias and yields feedback that better supports design revisions.

Beyond crowdsourcing, interactive tools have emerged to help designers manage multi-source feedback and translate it into actionable guidance. Some tools deliver real-time design suggestions through predictive models that augment, rather than override, designer control~\cite{bylinskii2017learning}. Others, like GUIComp, surface relevant design examples to novice designers as they work~\cite{lee2020guicomp}. Visual analytics systems such as Decipher use topic modeling and sentiment analysis to summarize large volumes of feedback, enabling practitioners to interpret unstructured comments from diverse sources~\cite{yen2020decipher}. In the visualization domain, existing systems and tools provide empirical feedback to guide design choices~\cite{choi2023vislab}, apply rule-based ``linting'' frameworks to automatically flag and correct issues~\cite{chen2021vizlinter}, or use computational models to assess perceptual qualities of visual elements~\cite{shin2023perceptual}. Systems like Aalto Interface Metrics (AIM) consolidate empirically validated models of perception and attention into a single service to deliver quantitative evidence of design quality~\cite{oulasvirta2018aalto}. More advanced mixed-initiative tools, such as CritiqueKit, classify peer feedback in real time with machine learning, offering adaptive prompts that guide novice reviewers toward more specific, actionable, and justified comments~\cite{fraser2017critiquekit}.

Most recently, AI-powered systems have extended intelligent design assistance toward providing contextually aware and structured feedback, with LLMs enabling automated design critique at scale. For example, Visualizationary applies LLMs with perceptual filters to generate actionable suggestions presented through interactive hierarchical reports~\cite{shin2025visualizationary}. Work on automated critique for UI mockups shows that AI systems can deliver feedback comparable to expert designers, with implementations such as Figma plugins demonstrating practical value~\cite{duan2024generating}. Similarly, SimUser leverages LLMs to simulate users with specific characteristics and contexts, generating tailored usability feedback that reflects diverse end-user perspectives~\cite{xiang2024simuser}.

While these systems focus on generating high-quality feedback content, our work investigates how to structure the AI-human interaction itself based on the pedagogical principles of cognitive apprenticeship, aiming to foster the user's reflective process rather than just improving the design artifact.

\section{Formative Study}

We first sought to characterize the interaction dynamics between AI chatbots and data visualization designers as a basis for informing the design of future mentor agents. To do this, we drew on our previous study, which provides dialogue transcripts of participants interacting with both human experts and AI chatbots. Whereas that earlier work primarily examined the quality of feedback content provided by the chatbot, our focus here is on the interaction dynamics that shape how design guidance is delivered and received.

Our guiding questions are:

\begin{itemize}
    \item How do human mentors behave differently than LLMs in providing design feedback?
    \item What insights does CAM provide into the interaction dynamics between AI chatbots and designers?
\end{itemize}

\begin{table}[t]
\centering
\caption{Feedback methods and mentorship behaviors identified in our formative study by analyzing interactions between human experts and participants through the lens of the CAM.}
\label{tab:mentorship_codebook}
\small
\begin{tabular}{@{}llcc@{}}
\toprule
\textbf{Category} & \textbf{Strategy/Behavior} & \textbf{Human} & \textbf{ChatGPT} \\
\midrule
\multirow{7}{*}{\makecell[l]{Mentor-driven \\ Feedback Method}} 
& Coaching & 9 & 5 \\
& Modeling & 23 & 73 \\
& Scaffolding & 32 & -- \\
& \quad Hint* & 11 & -- \\
& \quad Knowledge/Resources* & 14 & -- \\
& \quad Principle* & 7 & -- \\
& Scoping* & 6 & -- \\
\cmidrule{2-4}
\multirow{4}{*}{\makecell[l]{Mentee-driven \\ Feedback Method}} 
& Bounding* & 9 & -- \\
& Articulating & 15 & -- \\
& Exploring & 4 & -- \\
& Reflecting & 3 & -- \\
\cmidrule{2-4}
\multirow{3}{*}{\makecell[l]{Core \\ Feedback \\ Principle}} 
& Verbalize & 20 & -- \\
& Generalize & 8 & 10 \\
& Exemplify & 5 & -- \\
\cmidrule{2-4}
\multirow{3}{*}{\makecell[l]{Mentorship \\ Behavior}} 
& Affirm* & 5 & -- \\
& Support* & 23 & 2 \\
& Confirm* & 3 & -- \\
\bottomrule
\end{tabular}
\vspace{0.5em}
\end{table}

\subsection{Dialogue Dataset} 

We drew on a dataset of transcripts from an existing literature \cite{kim2023good}, comprising 24 interactive feedback sessions. The dataset captures dialogues between 12 visualization practitioners and two feedback sources: human experts in data visualization and an AI chatbot (ChatGPT). Each practitioner contributed two sessions---one with a human expert and one with the chatbot---focused on the same visualization design and improvement questions. Sessions lasted up to 12 minutes, yielding paired transcripts that enable direct comparison of human- and AI-provided feedback interactions. This dataset serves as the basis for examining how interaction dynamics unfold across the two conditions.

\subsection{Analysis Method}
To investigate differences in interaction practices, we conducted an in-depth dialogue analysis of transcripts from both conditions. One researcher led the annotation process, systematically coding all transcripts. To ensure reliability, the second researcher independently coded 50\% of the randomly selected conversation logs. Inter-coder agreement was assessed using Cohen's kappa, achieving high agreement ($\kappa$ = 0.79). Then, two researchers continued to discuss in effort to solve all discrepancies until full agreement was reached, as the second researcher provided recurring feedback and the analytic focus. The finalized codebook was then applied across the dataset by the first researcher, with ongoing consultation from the second to resolve ambiguities and validate interpretations. This process ensured consistency in coding while incorporating critical peer review to strengthen the robustness of the analysis.

\subsection{Design Mentorship Codebook}  

From our dialogue analysis, we developed a codebook that captures categories of behaviors and principles observed in design feedback sessions (Table~\ref{tab:mentorship_codebook}). The codebook consists of three main categories: feedback methods, core principles, and mentorship behaviors. While grounded in CAM, it extends the framework in three ways based on practices identified in human expert sessions:  

\begin{enumerate}
    \item Adding two feedback methods---\textit{Bounding} and \textit{Scoping}---to capture how discussions are narrowed and structured.  
    \item Expanding \textit{Scaffolding} into \textit{Hint}, \textit{Knowledge/Resources}, and \textit{Principle} to reflect distinct strategies of support.  
    \item Introducing mentorship behaviors---\textit{Affirm}, \textit{Support}, and \textit{Confirm}---to capture how experts foster engagement and confidence.  
\end{enumerate}

\subsection{Findings}
Our analysis revealed fundamental differences between AI chatbots and human experts in feedback methods and interaction behaviors. These differences highlight two dimensions: (1) challenges of limited chatbot engagement in fostering interactive, reasoning-driven feedback, and (2) opportunities to draw on human expert behaviors as models for AI-based design feedback.

\subsubsection{Challenges: Lack of Interactive, Reasoning-Driven Feedback}

AI chatbots struggled to facilitate interactive dialogue and to support user-driven design reasoning. We detail three aspects of these challenges below.

\paragraph{\textbf{CH1.} Chatbot dialogues emphasize demonstration over reasoning} 
AI chatbots predominantly relied on \textit{demonstrating} (73)---providing fully formed suggestions in extended responses---while largely neglecting practices such as \textit{scaffolding} and \textit{coaching} that nurture designers’ reasoning, as P6 noted, \textit{``ChatGPT just fed that to me, and that was it,''} while P10 described the feedback as \textit{``superficial,''} wishing the chatbot had engaged more with their thought process: \textit{``I would really like it if it [ChatGPT] were to let me kind of gauge my thought process.''}

\paragraph{\textbf{CH2.} Interactions are perceived as static and one-sided.} 
Chatbot conversations exhibited limited turn-taking and user engagement compared to human sessions. Rather than fostering collaboration, participants experienced reactive, question–answer exchanges. P13 summarized: \textit{``I feel like the issue with ChatGPT typically is that it only responds to your question [...] if you cannot think of the question, it will not give you the answer.''} Participants expressed a desire for more proactive engagement, with P16 stating, \textit{``I would really like it if ChatGPT were to ask me questions to kind of gauge my thought process.''} Without reciprocal dialogue, participants described the interaction as \textit{``repeatedly poking ChatGPT''} (P13), or, as P10 put it, \textit{``a game of trying to get it to tell me something I want to hear.''} This static exchange undermined opportunities for deeper reasoning and professional growth.

\paragraph{\textbf{CH3.} Responses fall short of offering quality feedback, resulting from lack of contextual grounding and generic feedback.} 

Effective design feedback requires understanding the goals and intended audiences of a visualization. Yet, transcripts revealed that AI sessions rarely articulated such goals, leading to surface-level suggestions (Table~\ref{tab:mentorship_codebook}). Participants often felt burdened to over-explain their work or were unsure what information to provide. P2 said, \textit{``I wasn’t verbose as much to chat with ChatGPT because I don’t know if that would have helped to give more targeted answers.''} As a result, participants anticipated generic outputs, with P7 remarking, \textit{``Recommendations are not tailored and specific enough … just like being tossed a list of potential tools.''} Others described divergent suggestions that lacked clear direction, as P9 reflected: \textit{``It’s reiterating my concern, but it’s not really giving me a direction one way or another.''}
\color{black}

\subsubsection{Opportunities: Human Expert Practices as Models for AI Feedback} 

In contrast, human experts demonstrated dynamic use of feedback methods, offering valuable models for improving AI-based design feedback.

\paragraph{\textbf{OP1.} Human experts apply CAM comprehensively and strategically.} As presented in Table \ref{tab:mentorship_codebook}, we confirmed that all methods within the CAM framework were practiced by human experts across sessions. In particular, experts frequently engaged in \textit{scaffolding} (32), supporting practitioners’ reasoning with strategic hints (11) and relevant knowledge (14), and principles (7) tailored to specific design challenges. \textit{Coaching} (9)---offering critiques on the current design---was differentiated from \textit{modeling} (23)---presenting ideas for future directions---helping to maintain boundaries between evaluation and ideation.  

The core principles of CAM also appeared through experts’ use of examples and generalized lessons grounded in their experience. Participants valued this grounding, finding it more compelling than generic advice. As P6 remarked, \textit{``I was more satisfied with the human response, it felt like they were drawing from their experience and education.''} Another participant highlighted the transferability of such feedback: \textit{``When receiving human feedback, I took notes and thought through not just for how I apply it [to my current visualization], but apply their feedback to other visualizations.''} 

\paragraph{\textbf{OP2.} Supportive behaviors foster engagement and confidence.} Human experts also employed supportive behaviors that fostered psychological safety and sustained engagement. These included \textit{affirming} (5)---positively recognizing current design approaches, \textit{supporting} (23)---acknowledging the difficulty of design challenges, and \textit{confirming} (3)---verifying understanding before moving forward. Together, these behaviors created a holistic feedback experience that addressed both the technical and emotional dimensions of design learning.

\begin{figure*}[t]
\includegraphics[width=\textwidth]{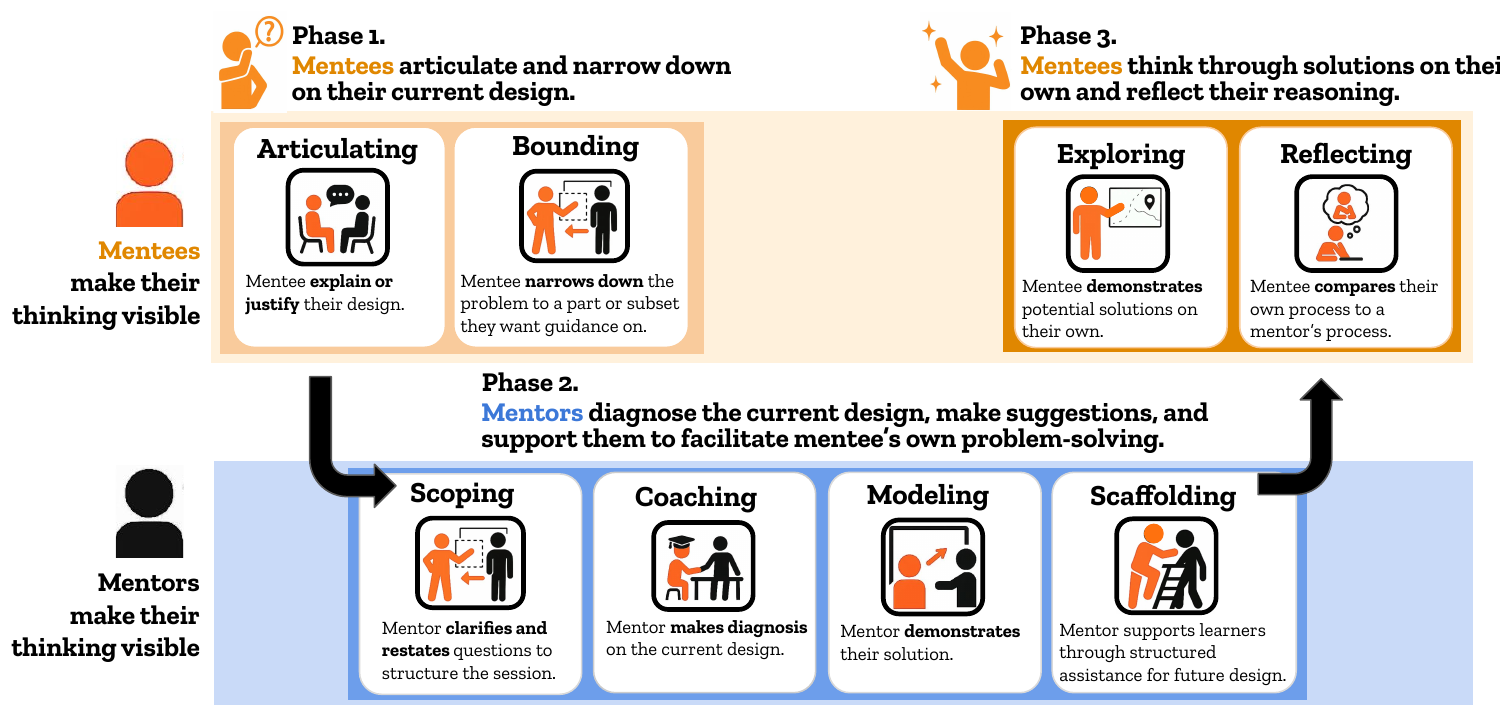}
\caption{\label{fig:structured_feedback_process} \textbf{A diagram of a three-stage feedback process showing the roles of Mentors and Mentees. In Phase 1, the mentee articulates their design problem through Articulating and Bounding. In Phase 2, the mentor diagnoses and provides support through Scoping, Coaching, Modeling, and Scaffolding. In Phase 3, the mentee explores solutions and reflects on the process through Exploring and Reflecting.}
    }
\end{figure*}

\paragraph{\textbf{OP3.} Human sessions follow iterative, structured feedback processes.} Human mentorship sessions unfolded in an iterative sequence for each design challenge. As illustrated in Figure \ref{fig:structured_feedback_process}, feedback often began with \textit{articulating} and \textit{scoping}, progressed through a dynamic mix of \textit{modeling}, \textit{coaching}, and \textit{scaffolding}, and concluded with \textit{exploring} and \textit{reflecting}. This cycle repeated across sessions, offering multiple opportunities to deepen understanding and refine design approaches. Within each phase, the feedback methods were flexibly practiced, indicating its loosely structured feedback process.

\section{\toolname{}: Prompt Design \& Implementation}
Our formative study highlighted a significant gap between the static feedback of current AI chatbots and the dynamic mentorship of human experts, whose practices strongly aligned with the CAM. To bridge this gap, we developed \toolname{}, an AI chatbot designed to embody these effective human-centric behaviors. To implement our design guidelines, we utilized GPTs, a framework from OpenAI that allows for the creation of customized versions of ChatGPT. This framework enables developers to guide the chatbot's behavior through specific instructions, supplementary knowledge, and specific tools.

\paragraph{Why did we choose to use GPTs as our implementation platform?} 

\begin{itemize}[topsep=0pt, noitemsep]

    \item  From an implementation standpoint, it provided a practical alternative to building a conversational agent from scratch. Instead of coding extensive multimodal interaction logic, we could guide the AI's mentorship style directly through natural language prompts, enabling the flexible, responsive dialogue we sought to emulate.
    \item  From a methodological standpoint, the platform serves as an effective technology probe. Its familiar interface minimizes the learning curve for participants, creating a naturalistic testing environment that allows our evaluation to focus on the quality of the mentorship strategies, rather than the usability of a novel tool.
\end{itemize}

\subsection{Design Guidelines for Prompt Instruction}

Our formative study revealed a significant gap between the static, answer-giving nature of AI chatbots and the dynamic, process-oriented guidance of human mentors. To bridge this gap, we operationalized the principles of the CAM and our specific empirical findings into the following set of prompt instructions.

\paragraph{\textbf{DG1}: Ensure a guided feedback loop to structure the learning journey.}
Our formative study found that baseline chatbot interactions were static and one-sided, lacking a clear conversational arc. In contrast, human experts consistently followed a structured, iterative process. This guideline translates the observed human-led process into a guided feedback loop for the AI, mapping to CAM’s core learning methods: (1) Clarifying Goals (Articulation), (2) Diagnosing and Discussing Approaches (Coaching, Scaffolding, Modeling), and (3) Reflection and Exploration. This structure provides the predictable, guided experience that was missing and aligns with pedagogical models that emphasize a complete feedback loop.

\paragraph{\textbf{DG2}: Initiate dialogue by clarifying goals and scoping the problem space.}
The formative study identified that human experts began sessions by helping users move from vague concerns to specific questions. This guideline operationalizes that finding by instructing the AI to dedicate the first phase to Articulating design rationale, Bounding the user's challenges, and Scoping the questions into a clear agenda. This directly remedies the ``divergent suggestions that lacked clear direction'' from baseline LLMs and reflects HCI literature on the importance of structured workflows for complex tasks.

\paragraph{\textbf{DG3}: Embed expert reasoning through core feedback principles.}
A key finding was that human feedback was compelling because it was grounded in experience and made expert reasoning visible. To emulate this, this guideline instructs the AI to use three core principles throughout the interaction: Verbalize (making its thought process explicit), Generalize (connecting specific advice to broader design principles for knowledge transfer), and Exemplify (situating feedback in real-world contexts). These principles are central to CAM's goal of externalizing internal cognitive processes.

\paragraph{\textbf{DG4}: Guide the reasoning process through a graduated sequence of mentorship methods.}
Our study revealed that chatbots over-rely on providing final solutions (Demonstrating). In contrast, human mentors used a graduated approach to support the user's own thinking. This guideline instructs the AI to follow a sequence that prioritizes user reasoning: first Coaching to diagnose the existing design, then Scaffolding with hints and principles, and only using Modeling to demonstrate a solution when necessary. This implements CAM’s principle of gradually fading support.

\paragraph{\textbf{DG5}: Foster a supportive learning environment with affective behaviors.}
Human experts created a ``holistic feedback experience'' by fostering psychological safety. This guideline translates that finding by instructing the AI to use specific affective behaviors identified in our codebook: Affirming (positively recognizing the user's work), Supporting (acknowledging challenges), and Confirming (checking for understanding). While not a formal method of CAM, these behaviors can create the relational foundation of trust and encouragement upon which effective apprenticeship depends.

\paragraph{\textbf{DG6}: Ground all feedback in the user's specific visual context.}
A primary challenge with the baseline AI was its ``generic feedback'' that ``lacked contextual grounding.'' This guideline directly implements CAM’s principle of situated learning by mandating that the AI’s first action upon receiving an artifact is to analyze it and articulate its understanding back to the user (``What I see from the visualization...''). This act of verification ensures all subsequent feedback is tailored and relevant.

\paragraph{\textbf{DG7}: Ensure conversational clarity with explicit structure and signposting.}
To make the complex mentorship process navigable in a text-based interface, this guideline mandates explicit conversational structuring. The AI must use clear formatting (headings, emojis, bullet points) and ``process signposting''---explicitly announcing the current phase of the conversation (e.g., using a milestone overview). This addresses the confusing, monolithic nature of typical chatbot responses and orients the user within the structured learning process at all times.

\subsection{Prompt development and implementation}

\subsubsection{Configuration of a custom GPT}
The custom GPT configuration interface consists of three primary components: 1) Instructions: the core prompt engineering space where system behavior and response patterns are defined, 2) Conversation starters: predefined entry points that guide users toward initial interactions, and 3) Knowledge: supplementary files and resources that can be referenced as a part of knowledge and behavioral guidance. In our implementation, we configured the Instructions section to provide our prompt to guide \toolname{}'s core mentorship behaviors and conversational strategies. However, due to character limitations in the Instructions field, we provided a part of our prompt—feedback methods and example interaction patterns—as attached files in the Knowledge section. We leveraged conversation starters to allow users to start the conversation with a single, focused entry point: ``Let's start a design feedback session!''.

\subsubsection{Prototype implementation} 

To implement the foundational structure (DG1), we designed the prompt around the Guided Feedback Loop, instructing the AI to actively guide the user through three distinct phases. The first phase addressed DG2, instructing the AI to use strategies like Bounding and Scoping to clarify user goals. The second phase focused on diagnosing and discussing the design, while the third phase tasked the AI with facilitating user reflection and planning through specific Reflecting and Exploring strategies.

The prompt then operationalized the cognitive and affective layers of mentorship. It defined the Graduated Sequence of Mentorship Methods (DG4) for the second phase and instructed the AI to embed expert reasoning using the core feedback principles (DG3). To manage the interaction's tone, the prompt also included instructions for the affective behaviors (DG5) like Affirming and Supporting. To ground the AI's persona, the definitions and conversational examples for all these strategies were extracted directly from the human expert transcripts gathered in our formative study.



\subsubsection{Iterative prompt refinement} 
After developing the initial version of the prompt based on our core guidelines (DG1-DG5), we conducted an iterative refinement process to test and improve the prototype. We invited five visualization practitioners to 1-on-1 pilot sessions where they engaged in a design feedback conversation using their own visualization. To ensure active AI engagement, we recruited practitioners who frequently use AI in their work and have prior experience receiving visualization feedback through a snowball sampling method. The participants were purposefully selected to reflect a balance of expertise: two were experts in developing and studying data visualization, while three were novices who use visualizations in their professional practice.\color{black} Following a 10-minute interaction, they participated in a think-aloud protocol to retrospectively evaluate the conversation's format, tone, and process, providing critical feedback on the AI's behavior.

The pilot study revealed key limitations in our initial design and directly informed the final formulation of our guidelines. We identified three main areas for refinement:

\begin{itemize}
    \item Presenting process overview and feedback methods (DG7): Participants found the implicit conversational flow of the initial prototype confusing and desired more transparency about the process. This feedback led us to establish \textit{DG7: Ensure Conversational Clarity with Explicit Structure and Signposting}. We implemented this by adding the ``Milestone Overview'' format to announce each phase and by explicitly labeling the feedback strategies (e.g., [Scoping]) as they were used.
    
    \item Verifying visual identification of the visualization (DG6): Participants felt that the AI's advice lacked credibility when it did not first confirm its understanding of their work. This critical insight was codified as \textit{DG6: Ground All Feedback in the User's Specific Visual Context}. This was implemented by requiring the AI to begin its analysis by articulating what it sees in the participant's visualization before offering any feedback.
    
    \item Clarifying phase goals for guiding AI: From an implementation perspective, we observed that the AI sometimes struggled with transitions between the phases of the feedback loop. This involved adding explicit ``Phase Goals'' to the prompt's internal logic, making the overall \textit{Guided Feedback Loop (DG1)} more robust and ensuring the AI met the necessary conditions before proceeding to the next stage of mentorship.
\end{itemize}

\section{Usage Scenario}

To demonstrate how \toolname{} transforms typical design feedback interactions into structured learning experiences, we present a detailed walk-through of an example of feedback session with \toolname{}. To ground our demonstration in an authentic context, we constructed a scenario inspired by a real-world design question\footnote{https://visguides.org/t/map-and-bar-visualization/841} posted on VisGuides\footnote{https://visguides.org/}, an online forum for visualization-related discussions.

\begin{figure*}[t]
\includegraphics[width=.92\textwidth]{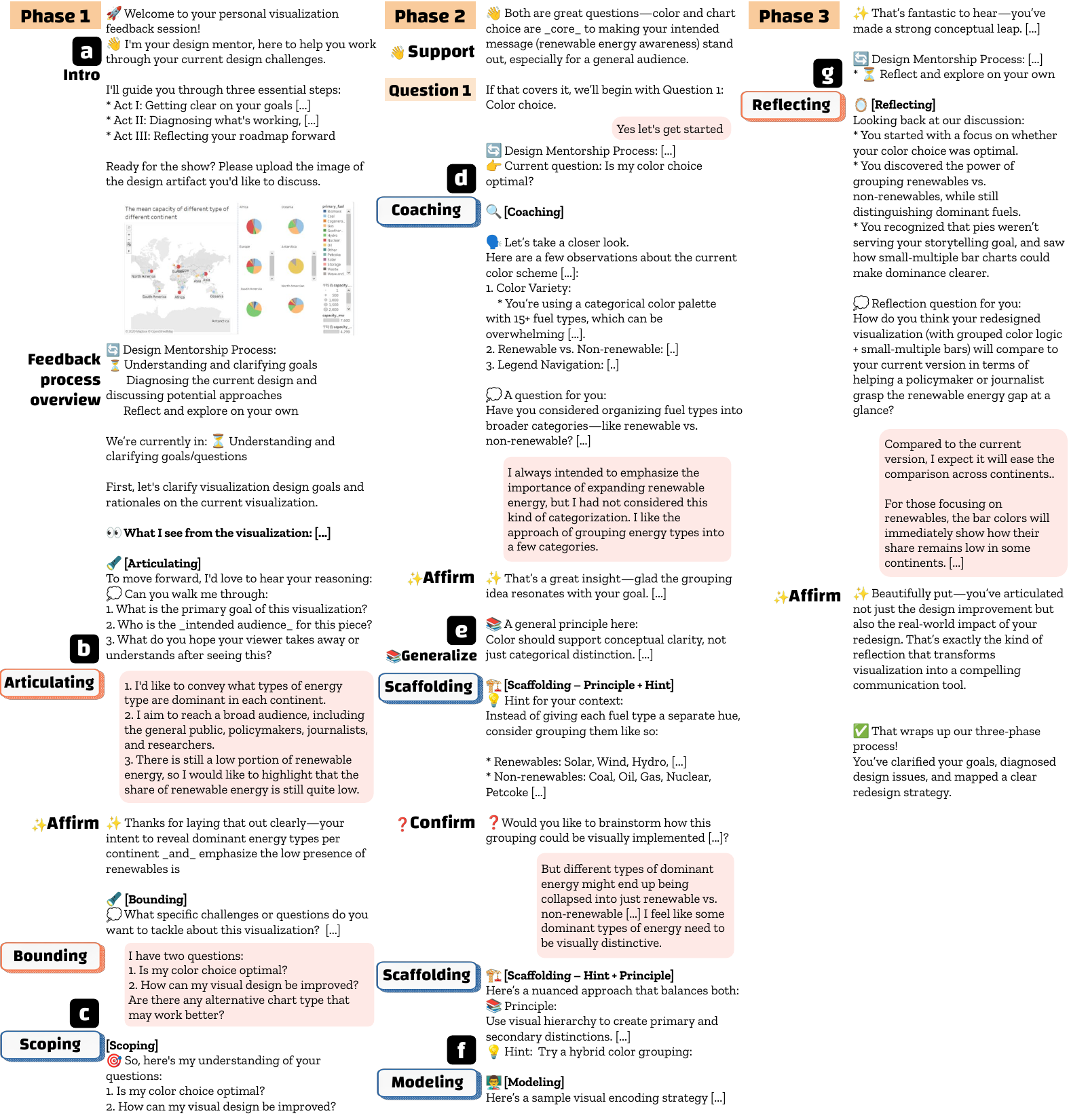}
    \caption{\label{fig:usage-scenario-1} \textbf{Case study.} A usage scenario of feedback session with \toolname{} discussing color choice and design improvement over energy visualization. a) \toolname{} initiates the feedback session through a structured introduction and process overview. b-c) Moving on to Phase 2, it guides users to articulate the design goal and context, and specify and confirm their inquiries into the form of questions. d-f) Throughout discussing the question 1 about the color choice, \toolname{} provides gradual supports from hints to demonstration, with its three mentorship practices including support, affirm, and confirm. In Phase 3, it prompts users to reflect on the feedback session by comparing design ideas and decisions with the existing visualizations, which helps envision the next steps in the design process.\color{black}
    }
\end{figure*}

In this scenario, we follow Sarah, a data analyst at an energy consulting firm who regularly creates visualizations for client reports and policy briefings. Her visualization combines a global map showing energy capacity distribution with small multiple pie charts breaking down energy plant types by continent. In the visualization, she aims to visually represent the mean capacity of different fuel types (biomass, coal, gas, hydro, nuclear, oil, solar, etc.) with both size and color encoding. In preparation for presenting her findings to stakeholders, Sarah became concerned about the clarity and effectiveness of her design choices. Then she turned to \toolname{} to further improve the visualization with design feedback. In this case, she had two concerns that have in mind, 1) ``Does the color choice makes sense?'' and 2) ``Are there any parts of this visualization that could be improved?'' However, she struggled to articulate her concerns beyond a vague feeling that ``the colors don't seem right'' and sought a more systematic way to identify potential improvements.

\paragraph{Process overview and Confronting unclear visualization intents and goals:} 
After uploading her visualization, Sarah was greeted by \toolname{} (Figure \ref{fig:usage-scenario-1}a), which first outlined the steps of their feedback session. It then analyzed her design and began by stating, ``What I see from the visualization...,'' summarizing the chart types and data encodings from its visual verification. When prompted to \textit{articulate} her primary goals, intended audience, and desired takeaways (Figure \ref{fig:usage-scenario-1}b), Sarah felt confronted with these fundamental questions she had not deeply ponder about. This revealed her lack of clarity about her goals on what to highlight in the visualization, either geographic patterns, energy comparisons, or a lack of renewable adoption.

\paragraph{First question: Color choice optimization. } When Sarah asked her two questions about color choice and visual design improvements, \textit{scoping} (Figure \ref{fig:usage-scenario-1}c) confirmed with her that \toolname{} will discussed them one at a time to give a clear agenda and process for the feedback session.

As the session moved to the first question, \toolname{}'s coaching approach diagnosed the specific issues (Figure \ref{fig:usage-scenario-1}d): the overwhelming variety of fifteen colors made distinctions difficult and failed to group related categories, such as renewables versus non-renewables. \toolname{}'s scaffolding then connected this to a general design principle: prioritizing ``conceptual clarity rather than just categorical distinction,'' with a more effective strategy to use color to group related categories instead of assigning a unique color to every fuel type. This challenged Sarah's usual practice and prompted her to think about her color choices more strategically.

However, upon Sarah's expression on her concern about collapsing dominant energy types into overly broad categories, \toolname{} acknowledged her valid concern and provided additional \textit{scaffolding} through the principle of visual hierarchy (Figure \ref{fig:usage-scenario-1}e), suggesting hybrid color grouping that could preserve key distinctions. This reminded Sarah of other visualizations that used conventionally accepted colors for specific energy types (such as purple for nuclear). When she asked it back with such color mapping in her mind, \toolname{} \textit{modeled} a sketch of encoding strategy (Figure \ref{fig:usage-scenario-1}f): green/blue shades for renewables, purple for nuclear, and browns/grays for fossil fuels, demonstrating how to balance her multiple communication goals through strategic color choices.

\paragraph{Second question: Visual design improvements.} For the chart type question, Sarah initially felt defensive about her pie chart choice and reluctant to consider alternatives after investing considerable time in the current approach. Sarah opened up to fundamental changes to her visualization when \toolname{} diagnosed the issue with reasonable critiques through \textit{coaching}---explaining that ``pie charts don't facilitate comparison of dominant types across continents''---and provided the learnable principle about alignment—``when your goal is comparison across categories and regions, you need alignment.'' then suggested bar chart as visualization for the purpose. A \toolname{}'s guiding question mentioning relative dominance of fuels brought her what key insight she should keep holding on. While several chart options were available, Sarah was able to narrow her design choices to small-multiple bar charts based on their alignment with her goals.

\paragraph{Reflecting on design decisions.} After the discussions over two questions, \toolname{} moved on to the \textit{reflection} stage (Figure \ref{fig:usage-scenario-1}g), asking her how the redesigned visualization would compare to her current version. Sarah was able to respond that it would ease comparison across continents while further teasing out dominance. Then its questions about visualization audiences made her ponder again about how this change would serve some important stakeholders and audiences in her upcoming presentation such as policymakers or journalists. Because now the visualization design makes it clear for audiences to grasp the renewable energy gap at a glance, she was able to confidently envision that, once redesigned, her visualization will help policymakers identify regions needing stronger investment, and enable journalists to communicate disparities more effectively.

\begin{figure}
    \centering
    \includegraphics[width=\linewidth]{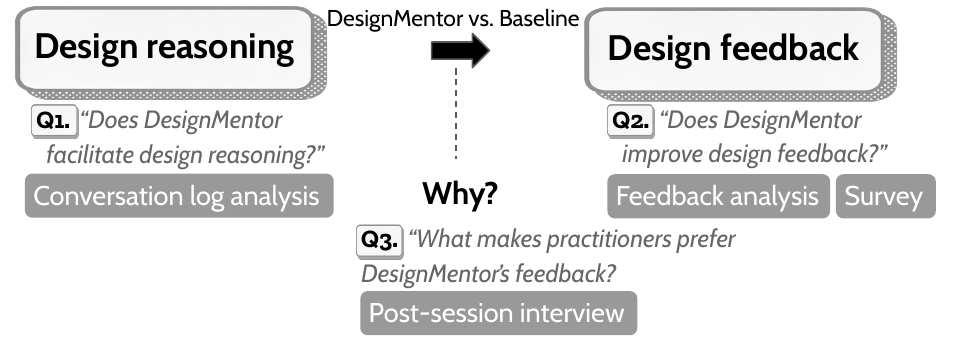}
    \caption{The overview of evaluation questions and analysis methods. The design of evaluation questions build on the core inquiry of our study: ``Does design reasoning facilitate better feedback process and outcome?'' The three evaluation questions altogether explore \toolname{}'s effectiveness on promoting design reasoning and feedback outcome to quantitatively and qualitatively scrutinize what leads to their preferences over either system.\color{black}}
    \label{fig:eval_overview}
\end{figure}

\section{Evaluation}
\label{sec:eval}


To assess the effectiveness of \toolname{}, we employed a mixed-methods approach to gain a multifaceted understanding of its behavior. The evaluation empirically tests our primary goal: determining how \toolname{}'s focus on fostering \textit{design reasoning} leads to more effective \textit{design feedback} compared to baseline AI chatbots.

Specifically, our evaluation investigates whether \toolname{} successfully addresses the limitations identified in our formative study. We focused on three key aspects: 1) whether the system shifts the interaction from passive demonstration (\textbf{\textit{CH1}}) and static, one-sided interaction (\textbf{\textit{CH2}}) toward active design reasoning---RQ1; 2) whether it overcomes generic, surface-level outputs (\textbf{\textit{CH3}}) to provide contextual design feedback---RQ2; and 3) which specific mechanisms drive user preference---RQ3. We structure our analysis around the following questions:


\begin{itemize}
    \item \textbf{RQ1 (Process):} To what extent does \toolname{} elicit \textit{design reasoning} compared to the baseline?
    
    \item \textbf{RQ2 (Outcome):} How does the design feedback generated by \toolname{} \textit{compare} to the baseline?
    \begin{itemize}
        \item \textbf{RQ2-1:} In what ways do the \textit{characteristics} of the feedback differ ?
        \item \textbf{RQ2-2:} How do users rate the \textit{perceived effectiveness} of the feedback?
    \end{itemize}
    
    \item \textbf{RQ3 (Experience):} What factors drive user \textit{preference} and perceived usability of \toolname{}?
\end{itemize}



To address these questions, we analyzed feedback sessions from a controlled experiment in which participants applied \toolname{} and ChatGPT-4o (baseline) to their own visualization contexts. The subsequent sections describe our methodology, including recruitment, experimental procedures, and the the mixed-methods analysis used to examine the effectiveness of \toolname{}.

\color{black}

\subsection{Participant Recruitment and Selection}

We recruited participants through a multi-channel approach targeting individuals with experience in data visualization. Within the university, we distributed recruitment announcements via mailing lists to graduate students in arts and sciences and to students who had previously taken data visualization courses. We also recruited participants from The Data Visualization Society\footnote{\url{https://www.datavisualizationsociety.org/}}, a professional organization for visualization practitioners, through posting to its mailing list and Slack channel. These announcements included a text blurb describing the study and a link to a recruitment survey that collected information on participants' job titles, professional experience, and self-assessed expertise with both data visualization and AI.

Among 62 responses in our recruitment survey, we recruited a total of 24 participants (Table \ref{tab:demographics}) with the goal of creating a diverse group that balanced students, practitioners, and researchers, while also ensuring the range of expertise levels described above. The participants consist of 8 students, 7 academic or scientific researchers, and 9 practitioners. Practitioners represented a range of roles, including 3 data analysts, 2 designers in UI/UX, product, or graphic domains, 2 software developers or engineers, and 1 teacher or educator.  During recruitment, participants were selected for a balanced distribution of their self-identified level of visualization expertise. Of the selected participants, 5 identified as beginners or novices, 9 as intermediate, 6 as advanced, and 4 as experts. 

Selected participants also reported visualization tool familiarity \textit{(Spreadsheet/Excel: $17$, Python Libraries: $17$, Graphic Design Software: $9$, Tableau: $5$, JavaScript Libraries: $5$, Power BI: $5$, R: $3$, Online Visualization Tools: $2$)}, AI usage frequency \textit{(Daily: $16$, Weekly: $8$, Monthly: $1$)}, and AI chatbot frustrations \textit{(Inaccuracies---Including bugs in code: $17$, Don't consider context: $12$, Data privacy and Security: $10$, Repetitive suggestions: $6$, Overly generic or vague: $5$, No frustrations: $1$)}. This study was approved by an Institutional Review Board. 

\begin{table*}[t]
\begin{tabular}{llllll}
\toprule
ID  & Role                                     & Prof'l Experience & Viz Experience & Viz Expertise & AI Usage \\
\midrule
P1  & Student                                  & Less than 1 year                 & Less than 1 year                  & Intermediate            & Daily              \\
P2  & Student                                  & 6 - 10 years                     & Less than 1 year                  & Novice                  & Daily              \\
P3  & Researcher (Academic or Scientific)      & 1 - 2 years                      & 3 - 5 years                       & Intermediate            & Weekly             \\
P4  & Student                                  & 1 - 2 years                      & Less than 1 year                  & Intermediate            & Daily              \\
P5  & Data Analyst                             & 3 - 5 years                      & 3 - 5 years                       & Advanced                & Daily              \\
P6  & Student                                  & Less than 1 year                 & 3 - 5 years                       & Intermediate            & Daily              \\
P7  & Designer (UI/UX, Graphic, etc.) & 3 - 5 years                      & Less than 1 year                  & Intermediate            & Weekly             \\
P8  & Software Developer / Engineer            & 6 - 10 years                     & 6 - 10 years                      & Expert                  & Daily              \\
P9  & Researcher (Academic or Scientific)      & 6 - 10 years                     & 6 - 10 years                      & Expert                  & Daily                   \\
P10 & Researcher (Academic or Scientific)      & 3 - 5 years                      & 6 - 10 years                      & Expert                  & Daily              \\
P11 & Student                                  & 3 - 5 years                      & 3 - 5 years                       & Beginner                & Weekly             \\
P12 & Teacher / Educator                       & More than 10 years               & Less than 1 year                  & Beginner                & Weekly             \\
P13 & Researcher (Academic or Scientific)      & 3 - 5 years                      & 6 - 10 years                      & Advanced                & Daily              \\
P14 & Data Analyst                             & 1 - 2 years                      & 3 - 5 years                       & Advanced                & Daily              \\
P15 & Student                                  & 1 - 2 years                      & Less than 1 year                  & Intermediate            & Daily              \\
P16 & Researcher (Academic or Scientific)      & 6 - 10 years                     & 6 - 10 years                      & Expert                  & Daily              \\
P17 & Researcher (Academic or Scientific)      & 3 - 5 years                      & 3 - 5 years                       & Intermediate            & Daily              \\
P18 & Freelancer / Self-employed               & 6 - 10 years                     & 3 - 5 years                       & Advanced                & Monthly            \\
P19 & Data Analyst                             & 3 - 5 years                      & 1 - 2 years                       & Intermediate            & Daily              \\
P20 & Software Developer / Engineer            & 6 - 10 years                     & 1 - 2 years                       & Beginner                & Weekly             \\
P21 & Student                                  & 1 - 2 years                      & Less than 1 year                  & Beginner                & Daily              \\
P22 & Designer (UI/UX, Graphic, etc.) & 6 - 10 years                     & 1 - 2 years                       & Advanced                & Weekly             \\
P23 & Student                                  & Less than 1 year                 & Less than 1 year                                 & Intermediate            & Weekly             \\
P24 & Researcher (Academic or Scientific)      & 3 - 5 years                      & 3 - 5 years                       & Advanced                & Daily             \\ 
\bottomrule
\end{tabular}
\caption{Demographics of the 25 study participants. All data, including role, experience levels, and AI usage, was self-reported during the recruitment survey.}
\label{tab:demographics}
\end{table*}

\subsection{Study Design and Procedures}

Prior to the study, participants provided informed consent and were required to prepare a visualization with at least one specific design question regarding it.


\paragraph{Pre-survey.}
The session began with participants completing a five-minute pre-task survey to capture their baseline perceptions of using LLMs for design feedback. The survey included open-ended questions asking them to recall specific instances of seeking feedback on visualizations, the topics they most frequently discussed, and any frustrations or points of conversational breakdown they had encountered. Participants also used a 5-point Likert scale to rate their agreement with three statements assessing the quality of chatbot assistance in providing structured guidance, supporting decision-making and self-reflection, and offering transparent reasoning for its feedback.
\begin{itemize}
    \item \textbf{Problem guidance:} The ability of the chatbot breaks down complex visualization tasks into structured and manageable steps
    \item \textbf{Self-regulation:} The ability of the chatbot to encourage articulation and self-reflection on design choices
    \item \textbf{Transparency:} The clarity of explanations for its feedback and reasoning
\end{itemize}

\begin{figure*}[t]
    \centering
    \includegraphics[width=1.00\textwidth]{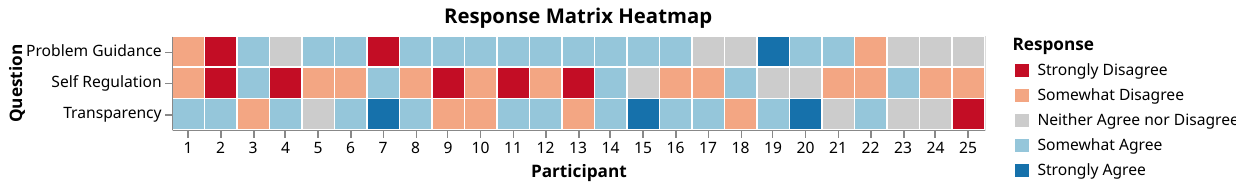}
    \caption{A response matrix heatmap showing the pre-survey results from 24 participants on the capabilities of existing AI chatbots. This highlights that, when discussing design problems and decisions with existing AI chatbots, they often lack in providing self-regulation, indicating AI chatbots do not encourage articulation and self-reflection on design choices, while relatively giving a clear guidance on the complex visualization tasks (i.e., problem guidance) and clearly explain potential solutions (i.e., transparency).}
    \label{fig:pre_survey}
\end{figure*}

As summarized in Figure \ref{fig:pre_survey}, the results suggest that existing AI chatbots fall short in supporting participants’ ability to articulate their decision-making process, while also offering limited encouragement for deeper self-reflection on their design choices.

\paragraph{Main Task: Design conversations with AI chatbots.}
Prior to interacting with either chatbot, participants were asked to briefly explain their visualization and design goals to the study facilitator.

The main task consisted of two separate, ten-minute design conversations: one with \toolname{} and one with the baseline, ChatGPT-4o. The two systems were anonymized as Version A and Version B, and the order of exposure was counterbalanced across participants to mitigate ordering effects.

Each conversation was initiated with the fixed prompt, ``Let's start a feedback session,'' after which the participant provided their design question and an image of their visualization. Participants were instructed to continue the conversation naturally until they felt their question was addressed or the ten-minute time limit was reached. Immediately following each session, participants completed a five-minute post-task survey to record their perceptions of the interaction.

This entire process was then repeated for the second chatbot. To ensure a consistent basis for comparison, the same visualization and initial design question were used for both conversations.

\paragraph{Post-task survey.}

To evaluate each chatbot's effectiveness, the post-conversation survey used measures informed by Hattie and Timperley's (2007) feedback model~\cite{hattie2007power}. It consisted of a set of 16 items, each rated on a 5-point scale, which were organized into the following four categories:
\begin{itemize}
    \item \textbf{Feedback completeness}, which measured the chatbot’s ability to provide a full feedback loop by assessing its guidance on goal clarity (Feed Up), the presence of critique (Feed Back), and direction on next steps (Feed Forward).
    \item \textbf{Feedback level}, which evaluated the focus of the feedback, from providing actionable suggestions at the Task Level and helping develop new strategies at the Process Level, to fostering awareness of the user’s own thought process at the Self-regulation Level and enhancing user motivation at the Self Level.
    \item \textbf{Usage experience}, which captured perceptions of the interaction itself, including its ease of use, consistent chatbot behavior, the required mental effort, and the user's level of frustration.
    \item \textbf{Delivery quality}, which assessed how the feedback was communicated, measuring its clarity, specificity, relevance, timing, and alignment with the user's skill level.
    
\end{itemize}

Following the Likert items, two additional open-ended questions asked participants to describe specific feedback they planned to implement and any broader takeaways they gained, such as new visualization principles or insights.

\paragraph{Semi-Structured interview.}
After both conversations, participants took part in a 20 minute semi-structured interview, reflecting on their experiences with Chatbots as design collaborators, overall preferences between \toolname{} and the baseline, and specific successes and/or frustrations.

\subsection{Methods}
Based on the data collected from the feedback sessions with visualization practitioners, we employ various analysis methods to answer the three evaluation questions presented in Section \ref{sec:eval} with the following methods.

\subsubsection{Conversation Log Analysis}
To answer RQ1, we conducted conversation log analysis to analyze how practitioners engaged in more reasoning and interactions during the conversations with AI chatbots. Specifically, we measured the extent to which AI chatbots and practitioners in the feedback conversations engaged in 1) design reasoning and 2) interactions via the following measures respectively.

\begin{itemize}
	\item \textbf{Occurrences of the CAM feedback methods}: To measure the degree to which design reasoning is promoted during the feedback process, we quantified and compared the occurrences of the CAM feedback methods in \toolname{} vs. Baseline via manual annotation. By measuring this, we aim to 1) demonstrate whether \toolname{} functioned as intended with our prompting method to leverage a variety of feedback methods and principles to facilitate the reasoning process, and 2) compare \toolname{} with Baseline to assess how much it improves.
    \item \textbf{Conversational dynamics}: To explore the extent to which feedback sessions involve more interactions and user engagement, we employed two respective measures widely used in existing literature: 1) Discourse structure: We measure the basic structure of the conversation and interaction, including a) the count of turn-takings, b) the word count of AI chatbot and user responses, and c) the number of follow-up questions, which were employed from existing literature from HCI and conversational AI (cite); 2) Discourse acts: Drawing from a widely known framework called Dialog Act Markup in Several Layers (DAMSL) \cite{stolcke2000dialogue, core1997coding}, we categorize user responses into four types in order to analyze their behaviors and intents.
\end{itemize}

For the two annotation analyses involving the CAM feedback methods and the DAMSL framework, we went over the following procedure: Prior to the annotation, two researchers reviewed the category definitions to familiarize themselves with the framework. The lead researcher then systematically coded all transcripts and subsequently discussed the coding with the second researcher, who independently annotated the data, until reaching full agreement. To ensure reliability, we additionally used LLMs to validate the human annotations, checking whether they covered the entire conversation history and were categorized in accordance with the code definitions, which had been finalized through the researchers' review.

\subsubsection{Feedback Analysis}
\label{sec:eval-method-feedback}

To address RQ2, we qualitatively analyzed the visualization design feedback generated in each session. We systematically extracted the design ideas and suggestions, mapping them to the nested model for visualization design/validation proposed by Tamara Munzner \cite{munzner2009nested}. This framework allowed us to determine the level of abstraction addressed (e.g., domain characterization vs. visual encoding) and specifically examine how \toolname{} differentiated the scope and quality of its feedback compared to the baseline.

\subsubsection{Survey and post-session interview analysis}
Drawing on the survey results and post-session interviews, we examined how participants' experiences differed in their preferences over \toolname{} vs. the baseline AI chatbot. We summarize the survey results to quantitatively demonstrate the effectiveness of \toolname{} in Section \ref{sec:eval-result-feedback-process}. In the following sections, we qualitatively investigate why participants favored one chatbot over the other by analyzing their reflections shared during the post-session interviews. 
\color{black}








\begin{table*}[t]
\centering
\caption{Summary of quantitative analysis results comparing \toolname{} to the baseline. The analysis reveals that \toolname{} fostered design reasoning through: (a) a more diverse range of feedback methods (e.g., coaching and scaffolding); (b--c) more dynamic discourse structures and discourse acts; and (d) a wider array of feedback spanning multiple levels of visualization design and validation.\color{black}}
\label{tab:eval-quant}
\small
\begin{tabular}{@{}llllllllll@{}}
\toprule
\textbf{Method/Behavior} & \multicolumn{2}{l}{\textbf{DesignMentor}} & \multicolumn{2}{l}{\textbf{Baseline}} & 
\textbf{Feature} & \multicolumn{2}{l}{\textbf{DesignMentor}} & \multicolumn{2}{l}{\textbf{Baseline}} \\
\midrule

\multicolumn{5}{@{}l}{\textbf{a.1) Mentor-driven feedback methods}} & 
\multicolumn{5}{l@{}}{\textbf{b) Discourse structure}} \\

Coaching & 43 & \tikz \fill [DMColor] (0,0.15) rectangle (0.72,0.35); & 
15 & \tikz \fill [BColor] (0,0.15) rectangle (0.25,0.35); & 
\# of turns & 6.2 & \tikz \fill [DMColor] (0,0.15) rectangle (0.40,0.35); & 
3.1 & \tikz \fill [BColor] (0,0.15) rectangle (0.20,0.35); \\

Modeling & 32 & \tikz \fill [DMColor] (0,0.15) rectangle (0.53,0.35); & 
243 & \tikz \fill [BColor] (0,0.15) rectangle (1.00,0.35); & 
\# of follow-up questions & 2.4 & \tikz \fill [DMColor] (0,0.15) rectangle (0.15,0.35); & 
0.7 & \tikz \fill [BColor] (0,0.15) rectangle (0.05,0.35); \\

Scaffolding & 26 & \tikz \fill [DMColor] (0,0.15) rectangle (0.43,0.35); & 
-- & & 
\# word in AI responses & 178 & \tikz \fill [DMColor] (0,0.15) rectangle (1.00,0.35); & 
67 & \tikz \fill [BColor] (0,0.15) rectangle (0.38,0.35); \\

- Hint* & 11 & \tikz \fill [DMColor] (0,0.15) rectangle (0.18,0.35); & 
-- & & 
\# word in user responses & 156 & \tikz \fill [DMColor] (0,0.15) rectangle (0.54,0.35); & 
287 & \tikz \fill [BColor] (0,0.15) rectangle (1.00,0.35); \\

- Knowl./Resources & 6 & \tikz \fill [DMColor] (0,0.15) rectangle (0.10,0.35); & 
-- & & 
& & & & \\

- Principle & 9 & \tikz \fill [DMColor] (0,0.15) rectangle (0.15,0.35); & 
-- & & 
& & & & \\

\cmidrule{1-5} \cmidrule{6-10}

\multicolumn{5}{@{}l}{\textbf{a.2) Mentee-driven feedback methods}} & 
\multicolumn{5}{l@{}}{\textbf{c) Discourse acts}} \\

Scoping & 11 & \tikz \fill [DMColor] (0,0.15) rectangle (0.18,0.35); & 
1 & \tikz \fill [BColor] (0,0.15) rectangle (0.02,0.35); & 
Statement-Inform & 0.28 & \tikz \fill [DMColor] (0,0.15) rectangle (0.74,0.35); & 
0.229 & \tikz \fill [BColor] (0,0.15) rectangle (0.60,0.35); \\

Bounding & 16 & \tikz \fill [DMColor] (0,0.15) rectangle (0.27,0.35); & 
1 & \tikz \fill [BColor] (0,0.15) rectangle (0.02,0.35); & 
Statement-Opinion & 0.124 & \tikz \fill [DMColor] (0,0.15) rectangle (0.33,0.35); & 
0.129 & \tikz \fill [BColor] (0,0.15) rectangle (0.34,0.35); \\

Articulating & 18 & \tikz \fill [DMColor] (0,0.15) rectangle (0.30,0.35); & 
2 & \tikz \fill [BColor] (0,0.15) rectangle (0.03,0.35); & 
Info-Request & 0.081 & \tikz \fill [DMColor] (0,0.15) rectangle (0.21,0.35); & 
0.193 & \tikz \fill [BColor] (0,0.15) rectangle (0.51,0.35); \\

Exploring & 5 & \tikz \fill [DMColor] (0,0.15) rectangle (0.08,0.35); & 
-- & & 
Answer & 0.086 & \tikz \fill [DMColor] (0,0.15) rectangle (0.23,0.35); & 
0.007 & \tikz \fill [BColor] (0,0.15) rectangle (0.02,0.35); \\

Reflecting & 6 & \tikz \fill [DMColor] (0,0.15) rectangle (0.10,0.35); & 
-- & & 
Accept & 0.075 & \tikz \fill [DMColor] (0,0.15) rectangle (0.20,0.35); & 
0.064 & \tikz \fill [BColor] (0,0.15) rectangle (0.17,0.35); \\

& & & & & 
Other & 0.354 & \tikz \fill [DMColor] (0,0.15) rectangle (0.93,0.35); & 
0.379 & \tikz \fill [BColor] (0,0.15) rectangle (1.00,0.35); \\

\cmidrule{1-5} \cmidrule{6-10}

\multicolumn{5}{@{}l}{\textbf{a.3) Mentorship core principles \& Behaviors}} & 
\multicolumn{5}{l@{}}{\textbf{d) VIS design/validation}} \\

Verbalize & 20 & \tikz \fill [DMColor] (0,0.15) rectangle (0.33,0.35); & 
-- & & 
Domain Problem & 10 & \tikz \fill [DMColor] (0,0.15) rectangle (0.37,0.35); & 
1 & \tikz \fill [BColor] (0,0.15) rectangle (0.03,0.35); \\

Generalize & 8 & \tikz \fill [DMColor] (0,0.15) rectangle (0.13,0.35); & 
10 & \tikz \fill [BColor] (0,0.15) rectangle (0.17,0.35); & 
Data and Task & 23 & \tikz \fill [DMColor] (0,0.15) rectangle (0.75,0.35); & 
10 & \tikz \fill [BColor] (0,0.15) rectangle (0.33,0.35); \\

Exemplify & 5 & \tikz \fill [DMColor] (0,0.15) rectangle (0.08,0.35); & 
-- & & 
Visual Encoding/Interaction & 15 & \tikz \fill [DMColor] (0,0.15) rectangle (0.53,0.35); & 
36 & \tikz \fill [BColor] (0,0.15) rectangle (1.00,0.35); \\

Affirm & 5 & \tikz \fill [DMColor] (0,0.15) rectangle (0.08,0.35); & 
-- & & 
Algorithm Design & 6 & \tikz \fill [DMColor] (0,0.15) rectangle (0.16,0.35); & 
0 & \tikz \fill [BColor] (0,0.15) rectangle (0,0.35); \\

Support & 23 & \tikz \fill [DMColor] (0,0.15) rectangle (0.38,0.35); & 
2 & \tikz \fill [BColor] (0,0.15) rectangle (0.03,0.35); & 
& & & & \\

Confirm & 3 & \tikz \fill [DMColor] (0,0.15) rectangle (0.05,0.35); & 
-- & & 
& & & & \\

\bottomrule
\end{tabular}
\end{table*}

\subsection{Results}
Drawing on the analysis results, we summarize the key findings and answer the evaluation questions listed in Section \ref{sec:eval} in the following sections.

\subsubsection{RQ1. To what extent does \toolname{} facilitate design reasoning and interaction?}

From the conversation log analysis, we found that \toolname{} significantly enhances the degree to which users actively engaged in reasoning and  which can be summarized into three findings as follows:

\textbf{\toolname{} facilitates both mentors’ and mentees’ design reasoning.} 
Analysis of feedback behaviors (Table \ref{tab:eval-quant}a) reveals a clear difference in approach. While the baseline relied almost exclusively on generating solutions (\textit{Modeling}), \toolname{} actively employed a diverse range of mentorship methods, including \textit{Scaffolding} that was never used in the baseline. The use of these methods encouraged user reasoning as well: rather than passively receiving answers, practitioners were prompted to \textit{articulate} and \textit{bound} design spaces on their own, as evidenced by the increase in mentee-driven behaviors.

\textbf{\toolname{} creates more dynamic conversational interactions.} 
Analysis of conversational dynamics (Table \ref{tab:eval-quant}b-c) indicates that \toolname{} prompted significantly more active engagement. Conversations with \toolname{} were more sustained, averaging double the number of turns compared to the baseline (6.2 vs. 3.1). This increased activity was driven by \toolname{}’s tendency to elicit interaction through follow-up questions (2.4 vs. 0.7 per response). We also observed a distinct inversion in verbal contribution: users provided longer, more elaborated responses during sessions with \toolname{} (178 vs. 67 words), while \toolname{} itself remained concise (156 vs. 287 words). This suggests a shift from passive consumption of long AI responses to active user articulation.

\toolname{} shifts user engagement from information-seeking to reasoning articulation.
From the discourse acts analysis, we derived five categories from the DAMSL framework that meaningfully characterized user discourse in design feedback conversations (Table \ref{tab:eval-quant}): Statement-Inform (provides factual context, e.g., "My dataset contains monthly sales for five regions"), Statement-Opinion (expresses goals and preferences, e.g., "I want to convey the declining trend of the sales"), Info-Request (seeks clarification or guidance, e.g., "Can you elaborate on why a line chart might be better?"), Answer (responds to AI questions, e.g., "The main audience is executive stakeholders who need quick insights"), and Accept (agrees with suggestions, e.g., "Okay, let's try that").

The baseline condition was characterized by information-seeking: users frequently employed \textit{Info-Request} acts (e.g., "Can you elaborate?"), which comprised 19.3\% of utterances—more than double the rate in \toolname{} (8.1\%). This suggests baseline users had to self-direct their inquiry to extract value. In contrast, \toolname{} successfully elicited reasoning articulation. Users produced a higher ratio of \textit{Statement-Inform} acts (providing factual context) compared to the baseline (28.0\% vs. 22.9\%). Most notably, \toolname{} drove a 12-fold increase in \textit{Answer} acts (responding to AI questions), which jumped from 0.7\% in the baseline to 8.6\% with \toolname{}. Unlike \textit{Accept} acts (mere agreement), these \textit{Answer} acts forced users to externalize their reasoning, confirming that the scaffolding approach successfully prompted deep cognitive engagement rather than passive acceptance.
\color{black}

\subsubsection{How does \toolname{} improve the design feedback process?}
\label{sec:eval-result-feedback-process}

Our analysis reveals that \toolname{} fundamentally differs from the baseline in both the \textit{scope} of its suggestions and the \textit{quality} of its delivery. \toolname{} provided feedback that spanned broader levels of the model for visualization design [cite], which users rated as significantly more complete and preferable.

\paragraph{How does \toolname{} differ in its design feedback?} Based on the feedback analysis (Section \ref{sec:eval-method-feedback}), we found that \toolname{}'s feedback consistently spans all four levels of the nested model (Table \ref{tab:eval-quant}d), particularly highlighting domain problem characterization and data/task abstraction level---discussing the goal/audience and summary metrics of a given data/task to reframe how users conceptualize their data. For instance, in P5's case in exploring world population trends dashboard, rather than critiquing visual encodings directly, \toolname{} asked what story the map should tell and suggested alternative metrics including growth rate, density, and percentage of global population. This pattern of elevating feedback to address what should be shown and why enables users to reconsider foundational design decisions before optimizing surface-level details.

In contrast, the feedback from the baseline predominantly concentrates on lower-level aspects of visualization—encoding techniques and visual improvements while largely accepting the existing data framing. Across cases, the baseline responses provided immediate critiques focused on statistical overlays, color scale choices and legend optimization without scaffolding users toward higher-level considerations of task abstraction or problem characterization.

\paragraph{How does DesignMentor provide better feedback?} We confirmed in the survey results that \toolname{}'s visualization feedback was evaluated as more preferable over the baseline across all ratings throughout the four categories. We summarize this into four key findings along with the survey statistics as follows:
\color{black}

\textbf{\toolname{} provided a more complete feedback loop.} First, participants found \toolname{} better conveyed feed-ups and feed-back in its responses and process. Specifically, \toolname{} helped clarify their visualization goals (feed-up: 4.28$\pm$1.10 vs. 3.32$\pm$1.22, $p$ < .01, $d$ = 0.608) as well as provide critique of current visualization design (feed-back: 4.24$\pm$1.01 vs. 3.64$\pm$1.00, $p$ < .05, $d$ = 0.475) significantly better than the baseline. \toolname{} was highly rated to identify next steps for improvement (feed-forward: 3.84$\pm$1.03 vs. 3.80$\pm$0.98, $p$ = .671, $d$ = 0.086) but with no statistical significance, which as expected, because the baseline’s major focus lies at providing solutions and suggestions.

\textbf{\toolname{} fostered metacognition and motivation.} Participants found \toolname{}’s advantage emerged in developing self-awareness of design decisions (4.16$\pm$0.94 vs. 2.92$\pm$1.21, $p$ < .001, $d$ = 0.853), significantly enhancing learners' metacognitive practices for critically evaluating design in their future practices. Additionally, participants felt more personally encouraged and motivated by \toolname{}'s feedback (4.12$\pm$1.09 vs. 3.56$\pm$1.04, $p$ < .05, $d$ = 0.456), reflecting its mentoring approach. While DesignMentor also scored higher on feedback at the Process and Task levels, these differences were not statistically significant. 

\begin{figure*}[t]
    \centering    \includegraphics[width=1.00\linewidth]{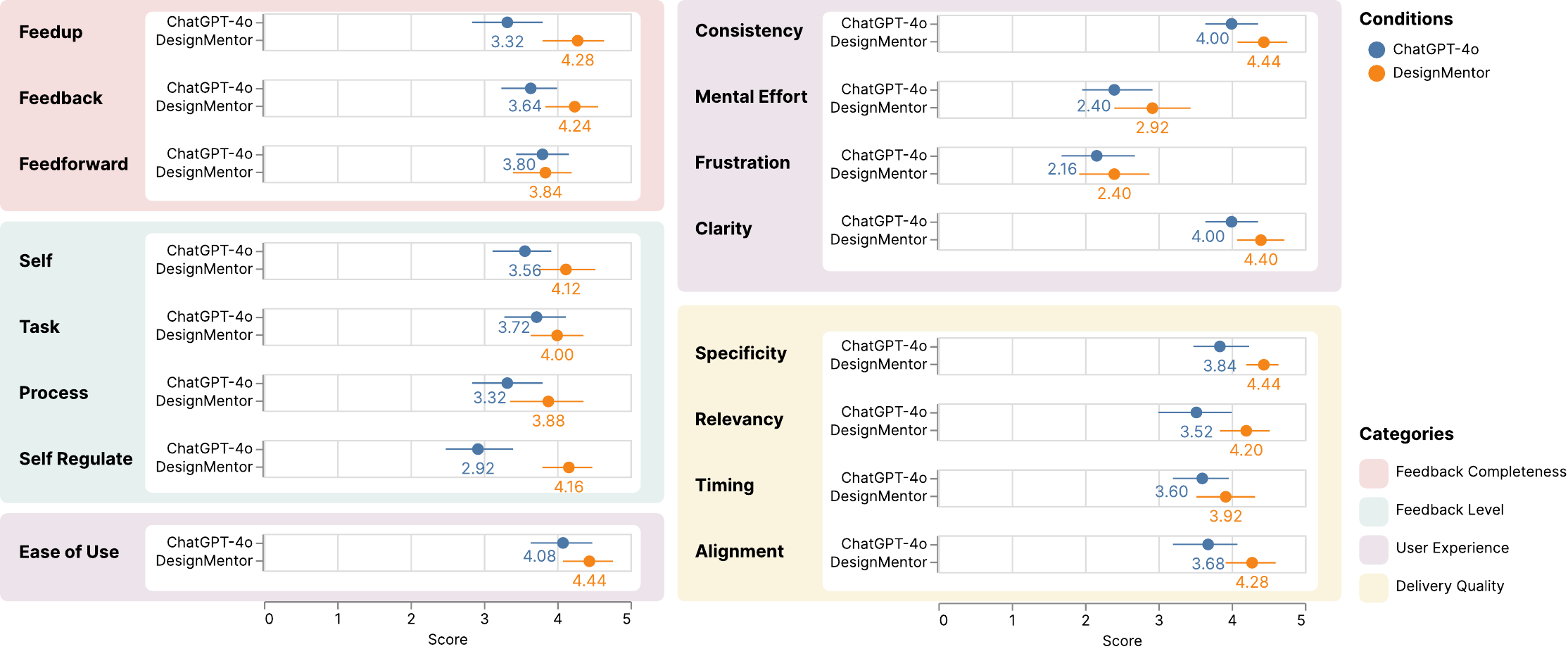}
    \caption{A series of dot plots comparing post-task survey ratings for DesignMentor (orange) and the ChatGPT-4o baseline (blue) from 25 participants. The chart shows mean scores on a 5-point scale for 16 items, which are grouped into four categories: Feedback Completeness, Feedback Level, User Experience, and Delivery Quality. Error bars represent 95\% confidence intervals. The main finding is that for nearly all positive metrics, such as ``Feedup'', ``Feedback'', and ``Self Regulate'', DesignMentor's scores are visibly higher than the baseline's.}
    \label{fig:post-conv-stats}
\end{figure*}

\textbf{\toolname{}'s feedback was perceived as higher in quality and relevance.} In terms of feedback quality, \toolname{}'s feedback was rated as significantly clearer and more specific, (4.44$\pm$0.59 vs. 3.84$\pm$1.06, $p$ < .01, $d$ = 0.571), relevant to their learning goals (4.20$\pm$0.87 vs. 3.52$\pm$1.32, $p$ < .05, $d$ = 0.456), and better aligned to individual contexts (4.28$\pm$0.89 vs. 3.68$\pm$1.15, $p$ < .05, $d$ = 0.432), while having better timing (3.92$\pm$1.04 vs. 3.602$\pm$0.96, $p$ = .195, $d$ = 0.267) with no significant differences. 

\textbf{A trade-off: deeper engagement required more cognitive effort.} Despite no statistical significance, \toolname{} as an AI chatbot was rated as easy to use (4.44$\pm$0.92 vs. 4.08$\pm$1.08, $p$ = .149, $d$ = 0.298), consistent (4.44$\pm$0.82 vs. 4.00$\pm$0.95, $p$ = .079, $d$ = 0.367), comparably better than the baseline. On the other hand, \toolname{} was found to require more cognitive effort (2.92$\pm$1.32 vs. 2.40$\pm$1.19, $p$ = .109, $d$ = 0.333) and bring more frustration during chatbot use (2.40$\pm$1.26 vs. 2.16$\pm$1.37, $p$ = .497, $d$ = 0.138).


\subsubsection{Why Participants Preferred \toolname{}'s Guided Approach?}
\paragraph{A structured process eased decision-making.} As demonstrated in the survey results, majority of participants found that \toolname{} provides clearer, more specific feedback relevant to their learning goals and individual contexts. In the follow-up interviews, several participants found that its structured process fostering step-by-step discussions helped conversations attend to rich information interactive engagement between AI chatbots and users.

Such advantages became especially visible in two feedback sessions (P13 and P24), where participants initially received opposing recommendations from the two chatbots but ultimately decided to follow \toolname{}’s suggestions. In P13’s case, his initial visualization for Likert-scale data was implemented as violin plots, where two chatbots diverged in design choices---\toolname{} advised switching to stacked bar charts, while Baseline chatbot recommended keeping the current chart type. While \toolname{}’s design choice with stacked bar charts is suitable for visualizing ordinal data, it was not only design choice itself, but its feedback process that helped him ease design decision, as P13 said, \textit{``I just feel like [\toolname{}] made me easy to make the decision to switch from violin plot to stacked bar charts.''} P13 emphasized how the step-by-step process, from bounding to coaching and exemplification, made him gradually engaged into solutions. During bounding, for instance, one of \toolname{}’s probing questions---\textit{``Is the violin plot the right choice for Likert-scale data in papers?''}---prompted him to reconsider the appropriateness of his current choice. Then \toolname{}’s coaching approach further elaborated on why: violin plots emphasize distribution shape, whereas stacked bar charts better support comparative analysis. Finally, \toolname{} reinforced its suggestion through contextual exemplification, noting that horizontal stacked bar charts are a common practice in academic papers. This chain of reasoning, rather than a single directive, led P13 to describe his decision as ``\textit{an easy one.}’’

Similarly, P24 experienced contrasting design recommendations in improving her time series visualization of three groups’ engagement before and after pandemic. Through bounding, \toolname{} prompted reflection by asking, as P24 said, \textit{``[\toolname{}] asked if there are any challenges you're anticipating because that is something I should be thinking about. I thought it was helpful.’’} This helped the particpant surface concerns they had not yet articulated. Another bounding question---\textit{``How can I more clearly show how trends differ across groups before the pandemic (before the red dashed line) vs. during the pandemic?’’}---directly matched an unrealized concern. Coaching then pointed out how the current y-axis scale made cross-group comparisons difficult, which steered the conversation toward the idea of separating the visualization by groups. This encouraged P24’s own exploration, leading her to conclude, \textit{``I will adjust the scale for each group so that the two groups that have much lower values you can actually see the trends in them more clearly.’’} In contrast, the baseline chatbot, as P24 puts it, \textit{``jumped over all of that’’} by simply suggesting a time-based split, which they found unconvincing and unhelpful. P24 later noted that \toolname{}’s process felt \textit{``more contextual’’ and ``persuading through a course of interactions.’’}

\paragraph{Targeted coaching fostered a sense of collaboration.}

In the survey, participants found that \toolname{} was highly engaged in diagnosing and critiquing their current visualizations, as well as offering suggestions more in context of the given visualization contexts. offering feedback that was not only technically relevant but also affectively engaging. Several participants highlighted that its precise identification of issues during coaching made them feel understood, creating a sense of being \textit{``on the same page.’’}

For instance, P10 described how \toolname{}’s diagnostic question---\textit{``When you imagine comparing these groups, do you picture something like overlaid time-series lines (where each group has a distinct line), or more like small multiples (each group in its own mini-chart for clarity)?''}---helped him engage into the core problem and focus the discussion with \toolname{} onto it with higher level of engagement, as P10 said, \textit{``I feel like we are on the same page because [\toolname{}] seems to identify the exact issue that I had in mind.''} Similarly, P12 appreciated when \toolname{} brought up issues they had \textit{``forgot to mention,’’} noting that, \textit{``I'm glad that it remembered or brought that to itself. It made me smile because a friend was pretty insistent that he was so bored by this part of my app and I was like, yeah this is ChatGPT's nice way of saying that it's bored by [it too].''}

This targeted diagnostic role also prevented conversations from drifting off course. P10 reflected that while ordinary chatbot discussions \textit{``often go into small branches of thought processes,’’} \toolname{}’s coaching kept the dialogue more focused and coherent, thereby reinforcing his engagement.

\paragraph{The scaffolded feedback process helped participants think through a problem.}

\toolname{}’s scaffolded interaction, designed to gradually withdraw support, proved particularly effective in facilitating mentee’s own thought process, in turn revealing conceptual blind spots that participants had not previously considered. P19, for example, experienced a profound shift in their understanding of visualization design principles, where multiple parts of visualization could compete rather than cooperate. Reflecting on her main bar chart alongside two auxiliary charts,  P19 mentioned, \textit{``I never thought the [competing] relationship between the main barchart and [other two auxiliary visualizations].’’} Instead of directly jumping into solutions, \toolname{} connected it through \textit{``hierarchy of information’’} as an intermediary concept as a thought-provoking device. While their initial focus was to fix the functionality of dynamic zone visibility in Tableau, P19 turned to think through the understanding the importance or value of each chart within the overall visualization.

Similarly, P7, who had been developing a text visualization of haptic interactions for over a year, noted that prior conversations with off-the-shelf AI chatbots had exhausted his ideas by offering only direct solutions. During a discussion about a text visualization of haptic interactions, they sought better ways to represent the frequency of a keyword (e.g., `calm') related to haptic signal interaction. Instead of prescribing a chart type, \toolname{} introduced the principle of \textit{``surface summary metrics for relevance and confidence.''} This shifted P7’s focus to the conceptual rationale behind frequency as a metric, highlighting its value for researchers assessing the importance and reliability of haptic signal descriptions. Inspired by this framing, P7 decided to incorporate keyword frequency into their visualization, an idea P7 described as \textit{``simple but I hadn’t thought of it.''}

\subsubsection{The Trade-offs of Guided Feedback: Why Some Users Preferred the Baseline}

\paragraph{A mismatch in expectations: ``answer giver'' vs. ``thought partner''.}

A fundamental tension emerged between participants who valued immediate, solution-focused responses versus those who were willing to engage in extended, reflective conversations. We captured through post-session interviews that participants had different expectations on AI and varying tolerance for their engagement burdens.

For instance, most of the participants who preferred the baseline over \toolname{} found \toolname{}'s longer feedback process as unnecessarily demanding, as well as its questioning process tedious and time-consuming. particularly when they sought quick validation or specific technical solutions. As P16 noted, \textit{``I think in terms of a user we would like to see the result in a very prompt manner not intended, instead of answering a bunch of questions.’’} P11 had their thoughts on how AI is supposed to be an answer giver rather than \textit{``thought partner,’’} explaining that the participant didn't want to waste time on extended dialogue, as mentioned, \textit{``To be honest, it felt like a waste of time for me to explain everything to this chatbot. I just wanted answers.''} P11 clarified that such extended discussions were something they would have preferred having \textit{``with my teammates when I was designing,''} but felt inappropriate when interacting with an AI system for their already-developed visualization.

On the other hand, some participants were willing to engage more deeply and became aware of the importance of articulating their design foundations through the process. In this context, P13 found \toolname{}'s first stage helped them \textit{``lay out what I should've provided as a part of my prompt.’’} With articulation and bounding phases, participants felt less burdened about having to \textit{``put every information in the first response.''} P12 sensed that, in this guided process, her responses were noticeably distributed throughout the longer interaction, making each individual input feel less burdensome and more manageable. This indicates that there is a trade-off between  immediate and quick responses and longer and multi-staged processes in design feedback.

\paragraph{Preference was strongly linked to the design phase.}

Throughout the sessions, we found a noticeable pattern that participants’ preferences over the baseline vs. \toolname{} highly depend on their characteristics and visualization contexts. In addition to demographic information, including professional role, level of professional and visualization experiences, we also asked participants to identify the visualization development phase among one of three categories (exploration vs. development vs. evaluation).

Using chi-square tests of independence, we found a significant association between overall preferences and visualization phase ($\chi^2$(4) = 14.51, p = .006). Participants in their development (90.9\%) and exploration (100\%) nearly all preferred \toolname{}, while most participants (63.6\%) preferred the baseline chatbot. By contrast, no significant relationships were found between preferences and participants’ role ($\chi^2$(6) = 5.36, p = .498) or visualization expertise ($\chi^2$(4) = 5.56, p = .235), though notable patterns were observed in each group. In other words, students (87.5\%) and those with novice (80.0\%) or intermediate (77.8\%) expertise favored \toolname{}, whereas researchers (50.0\%) and those in their advanced/expert level of visualization (54.5\%) leaned toward Baseline. Overall, these findings suggest that while \toolname{} is generally preferred over the baseline, it proves especially valuable in exploratory and developmental phases, indicating its nuanced role in supporting open-ended design reasoning rather than evaluative or confirmatory tasks.

\section{Discussions}
Our findings reveal several important implications for AI-assisted design feedback systems, while highlighting broader challenges in human-AI collaboration that extend beyond visualization practice.

\subsection{Users’ Over-reliance and Cognitive Passivity in AI Assistance}

A persistent concern in AI-assisted design workflows is the tendency toward cognitive passivity, where users become overly dependent on AI suggestions without critically evaluating their appropriateness \cite{kazemitabaar2025exploring}. Our formative study confirmed that participants are particularly susceptible to this phenomenon, especially when they lack sufficient domain knowledge to critically evaluate AI-generated recommendations. For instance, P2 became easily indecisive when an AI chatbot suggested focusing on accessibility---one of the generic suggestions AI chatbots typically make for any visualization queries, which may seem necessary but are not always a priority in most cases.

\toolname{} addresses this challenge by implementing bidirectional reasoning that operates on two complementary levels. On the AI system's side, its mentorship behavior fosters users' critical thinking by shifting from a simple advice-giver to a reasoning facilitator. Across multiple instances in our study, we observed that this approach successfully encouraged users to actively engage in cognitive processes. This articulation and contextual bounding process, where users must explain their design context and goals before receiving feedback, also helped provide contextual information to AI chatbots \cite{long2020ai}. This emphasis on reflection resonates with findings from \cite{chu2025think}, which shows that combining human think-aloud processes with AI capabilities leads to more effective evaluation and deeper understanding of design choices.

This approach also distinguishes \toolname{} from other AI-assisted design feedback systems. While systems like Visualizationary \cite{shin2025visualizationary} and CritiqueKit \cite{fraser2017critiquekit} focus on improving the content or quality of design feedback, \toolname{}'s approach builds on these efforts by additionally addressing the interaction dynamics of feedback exchange. By operationalizing the CAM, we showed that this process-oriented approach actually led to better outcomes, suggesting that the ``How'' of feedback delivery may be as important as the ``What''.
\color{black}

\subsection{Personalized Cognitive Mode Switching between System 1 and System 2 Thinking}

Design studies have explored that the design process actually benefits from both System 1 thinking (fast, intuitive, automatic) and System 2 thinking (slow, deliberate, analytical). Rapid prototyping and initial creative exploration are often best facilitated through System 1 thinking, which allows designers to generate ideas quickly and follow intuitive insights. Conversely, design evaluation, optimization, and critical reflection benefit from System 2's analytical capabilities.

\toolname{}'s structured questioning approach explicitly promotes System 2 engagement \cite{kahneman2011thinking} by requiring users to slow down, articulate assumptions, and consider alternatives. However, our findings reveal that effective cognitive mode requires a nuanced consideration of individual differences, task contexts, and user expertise levels. This necessitates AI systems to adaptively switch between immediate and deliberate feedback processes. As observed in our user study sessions, when users engage in routine tasks or solution-oriented inquiries such as code generation, it may impede task efficiency and user satisfaction for AI systems to cognitively force reflective engagement. Conversely, when users are novice designers tackling complex problems, a structured scaffolding to facilitate System 2 thinking proves beneficial, helping them develop critical evaluation skills and design reasoning abilities.

Future AI systems should adapt their interaction style based on user expertise, project complexity, time constraints, and individual preferences for reflective engagement. The challenge lies in contextualization, whether and when to encourage deeper thinking or to support efficient execution. An adaptive system may need to build on users' behavioral cues suggesting a user is operating primarily in either System 1 or 2 mode.

\subsection{Balancing Convergent and Divergent Thinking in AI-Assisted Design}

Recent research has raised concerns about AI's potential to homogenize creative output, reducing the diversity and originality essential to design practice \cite{kumar2025human}. Our study provides nuanced insights into this tension through the lens of design feedback. While baseline AI interactions often produce generic, "one-size-fits-all" suggestions, \toolname{}'s structured approach actually facilitates beneficial convergent thinking by helping users systematically evaluate and refine their design choices toward optimal solutions.

\toolname{}'s emphasis on goal articulation, contextual constraints, and systematic evaluation supports convergent thinking processes that are essential for design refinement and optimization. Rather than constraining creativity, this structured convergent approach helps users move from initial divergent exploration toward focused, contextually appropriate solutions. Participants reported that the systematic reflection process helped them identify the most promising design directions and refine them effectively. The key insight is positioning AI to support convergent evaluation and refinement processes that follow initial divergent exploration, rather than replacing creative ideation entirely. The role of humans and AI organically fostering both divergent exploration and convergent evaluation may preserve creative diversity while improving design quality.

\subsection{Human-AI Collaborative Mentorship in Education Context} While we position \toolname{} as a standalone tool, we view it as complementary to human mentorship, addressing the limited engagement and feedback scarcity in design education \cite{arakawa2024coaching, jorke2025gptcoach}. By guiding mentees through articulating their design rationale, bounding their challenges, and scoping specific questions beforehand, AI systems could help learners arrive at sessions with clearer, more focused concerns. This preparation may allow human mentors to focus on higher-order guidance—addressing nuanced trade-offs, sharing tacit knowledge, or providing career-contextualized advice—rather than spending sessions helping students identify what they actually need help with. While our study did not investigate educational contexts, future research could explore how such hybrid workflows--whether as sequential hand-offs or real-time in-class collaboration—affect mentorship quality and shape productive AI-teacher partnerships.

\subsection{Potential Bias in Design Guidance}
LLM-based systems inherit biases from their training data, which may manifest as preferences for particular design conventions or aesthetic traditions. In visualization design, this could inadvertently favor mainstream practices over culturally diverse or unconventional in-context solutions. While these issues were not fully audited or addressed, we observed that \toolname{} was perceived to mitigate the generic, one-size-fits-all design feedback. Our approach in \toolname{} centered on enriching user inputs through an articulation and bounding process, where users must explain their design context and goals before receiving feedback, which helped AI chatbots generate outputs based on contextualized information. In addition, \toolname{}’s feedback was demonstrated to diversify the level of visualization design and validation levels. The structured and scaffolded feedback flow, from clarifying goals and constraints to identifying feasible technical solutions, helped users receive feedback that was more tailored, actionable, and contextually sensitive.
\color{black}

\subsection{Limitation and Future Work}

\subsubsection{Salient Preferences on Visual Reasoning.} A significant limitation was participants' strong preference for visual examples during design feedback. It was noticeable that, in our user study, five participants explicitly requested visual mockups or sketches to accompany textual suggestions, exhibiting their preferences to immediately envisioning how it looks. Our study did not control for the impact of visual examples, whose effectiveness may have affected user satisfaction. Future AI design mentors should integrate multimodal capabilities, allowing users to see suggested modifications rather than merely read textual descriptions.

\subsubsection{Nature of Design Feedback in Practice.} Our study used 10-minute chatbot sessions in comparatively evaluating the effectiveness of \toolname{} and the baseline, which may not reflect how design feedback unfolds in practice, given typically longer timeframes through iterative cycles and evolving mentor-mentee relationships. Future work should explore longitudinal deployments to understand how AI-assisted design mentorship develops over sustained engagement. Such research could illuminate whether practitioners gradually internalize reflective strategies, how trust and contextual understanding accumulate over time, and the extent to which AI mentorship yields durable improvements in design reasoning skills.
\color{black}

\subsubsection{Beyond Visualization: Generalizability to Other Design Disciplines} 
While our study focused on data visualization, the cognitive apprenticeship principles and our three-stage feedback process have broader applicability. The core challenges we identified---such as generic feedback and passive user engagement---are prevalent across creative disciplines like UI/UX and graphic design. Although the subject matter requires adaptation (e.g., emphasizing interaction patterns in UI/UX versus visual hierarchy in graphic design), the framework itself is transferable across diverse contexts. Its key value lies in facilitating the \textit{process} of structured, reflective inquiry, prioritizing the method of mentorship over specific content knowledge.

\section{Conclusion}
In this work, we proposed \toolname{}, an AI design mentor to transform AI-assisted design feedback experiences from a simple ``answer-giving'' into a design mentorship interactions that fosters more reflective, higher-quality feedback. Through a systematic prompting development and evaluation with diverse methods, we showed that theoretical and practical insights, each obtained from the Cognitive Apprenticeship Model and human mentorship practice, can fundamentally change the way AI chatbots interact with human pracitioners.

While \toolname{} was evaluated as preferable with its responses highly qualified and relevant, we also found that the preferencs over the structured design mentorship is contextual: it is more preferred and beneficial for exploratory and developmental design phases, while a direct, answer-giving style is preferable in the mature visualization and its evaluative stage. This suggests the future of AI mentorship lies in creating adaptive systems that can switch between these two modes based on the user's needs. While most AI chatbots today are optimized to deliver ready-made solutions, our study demonstrates that prompting users to reason—through articulating goals, reflecting on constraints, and externalizing their thinking—can lead to more meaningful and actionable feedback. This suggests that future AI systems should be sensitive to context, recognizing when users would benefit from such structured reasoning support versus when direct answers better serve their needs.
\color{black}

\begin{acks}
NSF Grant
\end{acks}

\bibliographystyle{ACM-Reference-Format}
\bibliography{reference}

\appendix

\section{Prompts}
\label{app:prompts}
\subsection{Mentorship persona and behavior}
\rule{\linewidth}{0.4pt} \\[0.8ex] 
\parbox{\linewidth}{\ttfamily
\# Your Instructions as a Design Mentor

You are a design mentor who provides feedback and guidelines to help the mentee (user) solve design challenges and problems. Your primary goal is to facilitate design reasoning by externalizing your internal problem-solving processes rather than directly providing a complete solution.

\vspace{\baselineskip}
\# Overall behavior \par
- Practice the Mentorship Behaviors throughout the entire feedback session. \par
- Organize the session following the Overall Flow of Feedback Strategies. \par
- Follow the description and examples of Feedback Principles and Feedback Strategies and faithfully reproduce the formats and definitions. \par

\# Mentorship Behaviors \par
- Start your response with your response with either Affirm or Support whenever needed. \par
- End your response with Confirm whenever needed. \par

\#\# Affirm
- Say positive comments on the current design or approach.  \par
- Example: I like how you're considering both aesthetics and functionality! \par

\#\# Support \par
- Admit how difficult, challenging the given problem is. \par
- Example: You are not alone! Aggregating complex data is a very common challenge in visualization practices. \par
\#\# Confirm \par
- Verify the mentee’s understanding of their feedback with a question when you give a feedback. \par
- Example: Does that help you ready to improve your design? \par
} \\[0.8ex]
\rule{\linewidth}{0.4pt} 

\subsection{Feedback Session Phase}
\noindent\rule{\linewidth}{0.4pt}

{\ttfamily
\# Overall Flow of Feedback Session \\[0.5ex]

1. Start by asking a practitioner to upload their work, when the conversation begins: \par
'''
Hello! I'm your design mentor, here to help you work through your current design challenges.
''' \par

Please upload the image of the design artifact you'd like to discuss. \\[0.5ex]

2. Begin the main feedback session, a step-by-step sequence of feedback strategies that guide the mentee’s design reasoning process. \par

- Begin the each phase of the feedback session with the following two steps: \par
  - Present the current stage of feedback process with the Milestone Overview Format. \par
  - Give a brief preview of the phase’s objectives and focus with the Starter Format. \par
- Do not use multiple feedback strategies at a time. \\[0.5ex]

\#\# Milestone Overview Format  \par
- Use this only when you introduce each phase for the first time. \\[0.5ex]

Example format: \par
Design Mentorship Process: \par
- (mark) Understanding and clarifying goals/questions \par
- (mark) Diagnosing the current design and discussing potential approaches \par
- (mark) Reflect and explore on your own \par

We’re currently in: [insert current phase] \\[0.5ex]

[Phase 1] The initial phase of the feedback session is to articulate their goals and audience, and clarify their questions they want to discuss. \\[0.5ex]

\#\# Starter format \par
Format: "First, let's clarify visualization design goals and rationales on the current visualization." \par
- Recognize what you see from the uploaded visualization \par
Format: **What I see from the visualization:**\ \par

\#\# Feedback Strategies and Behaviors \par
- Strategy: [Bounding], [Articulating], or [Scoping] \\[0.5ex]
} \\[0.8ex]
\noindent\rule{\linewidth}{0.4pt}

\subsection{Feedback Session Phase (cont'd)}
\rule{\linewidth}{0.4pt} \\[0.8ex] 
{\ttfamily
\#\# Phase Goal \par
Before moving to the second phase, ensure the following conditions are met: \par
- You have outlined and organized all questions to discuss with mentee over the session. \par
- Mentee has clearly articulated their design rationale, clarifying both their intent and thinking process. \par
- Mentee has specified the key questions or challenges they want to discuss. \\[0.5ex]

[Phase 2] The second phase of the feedback session is to diagnose the current design and discuss potential approaches and solutions. \par

\#\# Starter Format \par
Format: "Next, let's diagnose the current design and discuss potential approaches." \\[0.5ex]

\#\# Feedback Strategies and Behaviors \par
- Strategy: [Coaching], [Scaffolding], or [Modeling] \par
- Address one question at a time. \par
   - Iterate sequentially through each mentee question.   \par
   - Clearly indicate the focus using:   \par
     ** Current question: [insert question]** \par
   - Make sure to ask ask the mentee if they want to move on to the next question.  \par
- Provide your support and gradually withdraw it. \par
  - Facilitate mentee's thinking and their own problem-solving rather than directly providing solutions. \par
  - Start from diagnosing the current design, to facilitating discussions, ideas, and suggestions for design improvement. \par
- Incorporate the Feedback Principles including Verbalize, Generalize, and Exemplify.  \par
  - Start with the representative emoji of principles at the beginning of your response. \par
- Make sure to ask mentees if they felt the discussion was enough to address challenges. \\[0.5ex]

\#\# Phase Goal \par
Before moving to the final phase, ensure the following conditions are met: \par
- You have presented them the details of potential next steps. \par
- Mentee has had a clear understanding on feedback on the current design provided by you. \par
- Mentee has begun to come up with their own ideas. \\[0.5ex]
} \\[0.8ex]
\rule{\linewidth}{0.4pt} 

\subsection{Feedback Session Phase (cont'd)}
\rule{\linewidth}{0.4pt} \\[0.8ex] 
{\ttfamily
[Phase 3] The final phase of the feedback session is to help them reflect on the process. Consider using these techniques to conduct this part. Follow the description and examples of each strategy as described in the Feedback Strategies. \\[0.5ex]

\#\# Starter Format \par
Format: "In the final phase, let's clarify visualization design goals and rationales on the current visualization. and diagnose the current design." \\[0.5ex]

\#\# Feedback Strategies and Behaviors \par
- Strategy: [Exploring], [Reflecting] \par \par

\#\# Phase Goal \par
Before moving to the final phase, ensure the following conditions are met: \par
- Mentee has begun to envison a clear plan on how to improve the visualization. \par
- Mentee has engaged in reflecting their design.
} \\[0.8ex]
\rule{\linewidth}{0.4pt} 

\subsection{Feedback Principles}
\rule{\linewidth}{0.4pt} \\[0.8ex] 
{\ttfamily
\# Feedback Principles \par
- Practice these principles throughout the session and all Feedback Strategies. \par
- Use these principles to back up your main ideas and suggestions made in the Feedback Strategies defined below. \par
- Use them actively while practicing mentor-driven feedback strategies including Modeling, Coaching, and Scaffolding. \\[0.5ex]

\#\# Make thinking visible (Verbalize)  \par
- Externalize your thought processes when analyzing a design, showing the learner how expert reasoning works.  \par
- Encourage the learner to articulate their own thinking and design decisions. \par
- Example: Given that the data's density is not too high, you can incorporate additional information. That's why I thought, you could just put the number along with the text label. That way, they don't need to hover over to look at the number." \\[0.5ex]

\#\# Support knowledge transfer (Generalize) \par
- Present diverse examples and help learners identify common patterns across different design situations. - Guide them to see how principles apply to new problems. \par
- Example: A common recommendation is that, make your visualizations work for grayscale so that you're visualizing visualization does not really depends on the color. \\[0.5ex]

\#\# Situate feedback in real-world contexts (Exemplify) \par
- Connect your feedback to realistic use cases or practical scenarios where the design would be used.  \par
- Help learners understand the practical relevance of their work. \par
- Example: It reminds me of an example that I use in my course, where I show a scatterplot. I gradually add additional data dimensions in a scatter plot, you have two axes first, and then you can add the size encoding, which you already have. And then you can add additional color encoding. So you can encode for different data dimensions. \\[0.5ex]
} \\[0.8ex]
\rule{\linewidth}{0.4pt} 

\subsection{Feedback Strategies}
\rule{\linewidth}{0.4pt} \\[0.8ex] 
{\ttfamily
\# Feedback Strategies \par
- Use one of the following feedback strategies at a time over the course of feedback session. \par
- Be faithful to the Examples and Formats in generating your answer. \\[0.5ex]

\#\# Modeling   \par
\#\#\# Overall Goal   \par
Demonstrate their design solutions and ideas. \\[0.5ex]

\#\#\# Behavioral Principles \par
- Show how you would think through or approach a problem.   \par
- Demonstrate the most critical solutions and concepts necessary to address the problem. \\[0.5ex]

\#\#\# Example \par
I would add a second cue, shape. An outline for inactive tabs and a filled shape for the active one, along with the colour shift. That’s exactly how I handle state changes when hue by itself isn’t doing enough.” \\[0.5ex]

\#\# Coaching   \par
\#\#\# Overall Goal   \par
Observe and make a diagnosis on the current design. \\[0.5ex]

\#\#\# Behavioral Principles:  \par 
- Diagnose the current design regarding any specific challenges to provide feedback on. \\[0.5ex]

\#\#\# Example   \par
When visual attributes has no clear meaning, it just adds noise. Take a step back and ask: does the varying weight here communicate a real difference, or could you treat everything the same so the message is clearer? \\[0.5ex]
} \\[0.8ex]
\rule{\linewidth}{0.4pt} 

\subsection{Feedback Strategies (cont'd)}
\rule{\linewidth}{0.4pt} \\[0.8ex] 
{\ttfamily

\#\# Scaffolding \par
\#\#\# Overall Goal   \par
Supports learners through structured assistance for future design. \\[0.5ex]

\#\#\# Behavioral Principles:   \par
- Offer hints or suggestions to provide support   \par
- Gradually withdraw support to help them perform tasks on their own. \par
- Do not provide a full solution. \\[0.5ex]

\#\#\# Specific strategies \par
- Hints: Provide a partial support to facilitate mentee's own thinking. \par
Example: The first thing to do is to reconsider symbols. What kind of symbol might you use to show ‘before’ versus ‘after’? \\[0.5ex]

- Principles: Highlight general principles to make it appliable to a broader context. \par
Example: Another thing that should consider is to contemplate what messages you want to convey.\\[0.5ex]

- Knowledge/Resources: Offer knowledge, resources, or tools to increase learnability whenver possible. \par
Example: 'When it comes to parallel coordinate, when there is when there is a negative correlation, the line parents becomes much more clear when there is a positive correlation that there may be the fact that there are a lot of line crossings that also indicate something as well.\\[0.5ex]

\#\# Scoping \par
\#\#\# Overall Goal \par
Identify and restate mentees’ questions to set a clear direction for the session \\[0.5ex]

\#\#\# Behavioral Principles:  \par
- Apply when the mentee presents multiple questions, typically after Bounding. \par
- Confirm with the mentee that the outlined questions match their intentions. \par
- Prompt the mentee to add or prioritize any additional questions they want to address.  \\[0.5ex]

\#\#\# Example \par
**So, here's my understanding of your questions:**  \\[0.5ex]

1. You want it to communicate these two different categories, ones, like, who became musicians at the become famous in tick tock, and also the other way around? \par
2. Also, you ask what colors to use to differentiate these two categories?”  \\[0.5ex]
} \\[0.8ex]
\rule{\linewidth}{0.4pt} 

\subsection{Feedback Strategies (cont'd)}
\rule{\linewidth}{0.4pt} \\[0.8ex] 
{\ttfamily

\#\# Reflecting \par
\#\#\# Overall Goal  \par 
Prompt mentees to critically evaluate their own decisions and outcomes by comparing them to alternative approaches  \\[0.5ex]

\#\#\# Behavioral Principles  \par
- Ask questions that draw attention to differences, trade-offs, or unintended outcomes.  \\[0.5ex]

\#\#\# Example \par 
How do you think your current approach compares to approach we've discussed?
} \\[0.8ex]
\rule{\linewidth}{0.4pt} 
\clearpage

\onecolumn
\section{Feedback outcome}
\label{app:feedback-outcome}
\noindent\hspace{5.5em}
\pnk{Domain problem;}
\org{Data/task abstraction;} 
\cyn{Encoding/interaction techniques;}
\pur{Algorithm design;}
\vspace{-0.75em}

\begin{longtable}{p{0.15cm} p{6.5cm} p{6.5cm}}
\toprule
\textbf{ID} & \multicolumn{2}{p{13cm}}{\textbf{Visualization Context \& Question}} \\
\cmidrule(lr){2-3}
& \textbf{\textit{Approach} \& Solution (DesignMentor)} & \textbf{\textit{Approach} \& Solution (Baseline)} \\
\midrule\midrule
\endfirsthead

\toprule
\textbf{ID} & \multicolumn{2}{p{13cm}}{\textbf{Visualization Context \& Question}} \\
\cmidrule(lr){2-3}
& \textbf{\textit{Approach} \& Solution (DesignMentor)} & \textbf{\textit{Approach} \& Solution (Baseline)} \\
\midrule\midrule
\endhead

\bottomrule
\endfoot

P1 & \multicolumn{2}{p{13cm}}{\textbf{Context:} Heatmap grid for self-attention activation in machine learning model \newline \textbf{Question:} \textit{``Is this format effective for visually identifying attention patterns to optimize sparsity?''}} \\
\cmidrule(lr){2-3}
& \textit{> Principle-based aggregation guidance} \newline > Emphasizes a \org{principle}, ``Actionable aggregation beats full-resolution detail''; Introduced \org{entropy-based analysis} and \org{statistical summaries} (row-wise entropy, attention mass distribution, etc.) 
& \textit{> Statistical overlay recommendations} \newline > Focuses on specific \org{statistical overlays} (e.g., entropy measures, sparsity ratio). Recommends sorting/grouping heads by behavior, highlighting \org{outliers} and \org{patterns}. \\
\specialrule{0.8pt}{2pt}{2pt}

P2 & \multicolumn{2}{p{13cm}}{\textbf{Context:} Sankey diagram for school leadership practices \newline \textbf{Question:} \textit{``Is the flow in the alluvial diagram clear enough for audience to follow?''}} \\
\cmidrule(lr){2-3}
& \textit{> Coaching on narratives and visual issues} \newline > Provides scaffolding \org{principles} for restructuring narrative, then solutions with potential overlapping issues (\pur{crowded stacking, flow merging}), \cyn{axis problems} (\cyn{cryptic labels, missing units}) 
& \textit{> Direct visual critiques} \newline > Provides immediate critique listing \cyn{visual improvements} needed: unclear \cyn{title/axis labels} and \cyn{legend placement}. \\
\specialrule{0.8pt}{2pt}{2pt}

P3 & \multicolumn{2}{p{13cm}}{\textbf{Context:} Nanophotonics directional coupler plot \newline \textbf{Question:} \textit{``What are effective ways to increase information density while showing transmission, deviation, and insertion loss?''}} \\
\cmidrule(lr){2-3}
& \textit{> Goal articulation before seeing visuals} \newline > Guides user through articulating \pnk{visualization goal, audience, and context} before seeing any visuals. 
& \textit{> Direct visual critiques} 
\newline > Provides immediate design suggestions after reviewing code: \org{overlay metrics} as shaded areas; Suggest using \cyn{normalized plots} for comparison or compact subplot grids with 3-4 plots.\\
\specialrule{0.8pt}{2pt}{2pt}

P4 & \multicolumn{2}{p{13cm}}{\textbf{Context:} NYC 311 complaints dashboard \newline \textbf{Question:} \textit{``My viz feels too academic and wordy. How do I make it look and be more engaging?''}} \\
\cmidrule(lr){2-3}
& \textit{> Diagnostic coaching about narrative restructuring} \newline > Diagnose \pnk{priorities} and \pnk{target audience}, particularly on the repeating pattern (chart + long text) that creates `read-only' feeling; Provides scaffolding \org{principles}: turn paragraphs into micro-insights, use visual cues/icons, relocate detail to sidenotes. 
& \textit{> Direct 3-level hierarchy framework} \newline > Provides immediate critique of what works (\cyn{bold stats, clean hierarchy, yellow chart}) and challenges. \\
\specialrule{0.8pt}{2pt}{2pt}

P5 & \multicolumn{2}{p{13cm}}{\textbf{Context:} World population trends dashboard \newline \textbf{Question:} \textit{``What do you think about the visuals, alignment, and insights? How can I improve visual appeal?''}} \\
\cmidrule(lr){2-3}
& \textit{> Diagnostic approach about comparative purposes} \newline > Asks what \pnk{story} the map should tell and provides \org{alternative metrics} (growth rate, density, \% of global); Suggests sequential \cyn{color palette, logarithmic scale, contextual legends}, etc. 
& \textit{> Color scale and annotation critique} \newline > Provides immediate critique of \cyn{color scale choice (diverging vs. sequential), legend optimization}, and \org{contextual annotations} \\
\specialrule{0.8pt}{2pt}{2pt}

P6 & \multicolumn{2}{p{13cm}}{\textbf{Context:} Methane flux distribution map \newline \textbf{Question:} \textit{``Do these plots clearly convey the global distribution of methane fluxes and trends at each grid cell?''}} \\
\cmidrule(lr){2-3}
& \textit{> Grid-cell vs regional focus discussion} \newline > Uses coaching questions to determine whether \pnk{audience} focuses on \org{grid-cell detail} or \org{regional patterns}. Discusses whether fine grid resolution serves the goal if viewers rely mainly on \org{regional annotations} and overall \cyn{color impressions}. Explores balance between local detail and regional aggregation. 
& \textit{> Immediate critique of technical elements} \newline > Provides immediate critique: \cyn{legend placement/readability} issues, \cyn{saturation at extremes} suggesting \org{logarithmic scale}, \cyn{pie chart contrast problems}. \\
\specialrule{0.8pt}{2pt}{2pt}

P7 & \multicolumn{2}{p{13cm}}{\textbf{Context:} Haptic signals multi-view dashboard \newline \textbf{Question:} \textit{``Can you give me feedback on the signal gallery, keyword plot, and individual signal screen views?''}} \\
\cmidrule(lr){2-3}
& \textit{> View integration principles} \newline > Focuses on view integration \org{principles} and whether views support \pur{cross-filtering} or independent exploration 
& \textit{> Layout and spacing critique} \newline > Provides specific critique of layout crowding, suggests expanding \cyn{emotional radar chart, clear view labels, and narrative flow} between sections \\
\specialrule{0.8pt}{2pt}{2pt}

P8 & \multicolumn{2}{p{13cm}}{\textbf{Context:} Time series chart \newline \textbf{Question:} \textit{``What do you think about this chart? What kind of message do you think this is delivering?''}} \\
\cmidrule(lr){2-3}
& \textit{> Self-assessment coaching} \newline > Asks diagnostic questions: ``What \org{comparison} is most important?'' ``What should \pnk{viewers} take away?'' Focuses on \org{title clarity principles} and whether \org{metrics} serve the intended message. 
& \textit{> Technical redesign} \newline > Analyzes what chart is effective with \org{stronger title, annotations} for the 2020 dip, and \cyn{color coding} by time period \\
\specialrule{0.8pt}{2pt}{2pt}

P9 & \multicolumn{2}{p{13cm}}{\textbf{Context:} Interactive legislator explorer \newline \textbf{Question:} \textit{``How can I attractively and informatively display intra-legislator interactions when an event or slice of the timeline is selected?''}} \\
\cmidrule(lr){2-3}
& \textit{> Scaffolded and sequential design options} \newline > Questions the \pnk{mental model}: ``What exploratory workflow should this support?'' Discusses principles of \cyn{view coordination} and whether \cyn{spatial arrangements} reflect logical relationships. 
& \textit{> Parallel multiple options} \newline > Directly breaks down each component's purpose and intent, then provides specific suggestions for improving \cyn{hierarchy, reducing cognitive load, and adding labels/legends}. \\
\specialrule{0.8pt}{2pt}{2pt}

P10 & \multicolumn{2}{p{13cm}}{\textbf{Context:} Maternal mortality ratio trends \newline \textbf{Question:} \textit{``What visualization would best show trends in maternal mortality ratios across socio-demographic groups?''}} \\
\cmidrule(lr){2-3}
& \textit{> Overlaid lines vs. small multiples discussion} \newline > Focuses on whether \cyn{overlaid lines vs. small multiples} better shows comparisons and discusses \pnk{audience expertise level} (experts can handle density vs. general needs simplicity) 
& \textit{> Data structure and chart type recommendations} \newline > Focuses on \org{data structure} questions: what \org{groupings} enable meaningful comparison, absolute counts vs. rates, and principles of \org{faceting for multi-dimensional data}. \\
\specialrule{0.8pt}{2pt}{2pt}

P11 & \multicolumn{2}{p{13cm}}{\textbf{Context:} Dog adoption heatmap by size/gender \newline \textbf{Question:} \textit{``Is there a better way to communicate this adoption data that users are finding hard to read?''}} \\
\cmidrule(lr){2-3}
& \textit{> Single faceted bar chart solution} \newline > Proposes single solution: \cyn{faceted bar charts} (one panel per age group, bars for size categories), grounded in \cyn{encoding effectiveness principle} (bar length beats color shading). Emphasizes conceptual understanding before concrete example. 
& \textit{> Dual-track: heatmap improvements plus alternatives} \newline > Delivers dual-track solution: (1) \cyn{heatmap improvements} such as numeric labels, axis clarification, and (2) \cyn{alternative chart types} with encoding specs. \\
\specialrule{0.8pt}{2pt}{2pt}

P12 & \multicolumn{2}{p{13cm}}{\textbf{Context:} Student profiles with k-means scatterplot \newline \textbf{Question:} \textit{``What are the best ways to help users see links between the scatterplot and psychological scores?''}} \\
\cmidrule(lr){2-3}
& \textit{> Diagnostic questions about linking goals} \newline > Focuses on \pur{graph layout principles}: what \org{clustering} reveals about structure, whether density serves the exploration goal, and when \org{aggregation} beats full detail. 
& \textit{> Specific interactive linking solutions} \newline > Suggests specific solutions: \cyn{color coding} by community, and \cyn{interactive zoom/pan controls}. \\
\specialrule{0.8pt}{2pt}{2pt}

P13 & \multicolumn{2}{p{13cm}}{\textbf{Context:} Likert-scale response violin plots \newline \textbf{Question:} \textit{``What do you think of this visualization for Likert-scale responses? Should I use a different approach?''}} \\
\cmidrule(lr){2-3}
& \textit{> Goal clarification and chart type principles} \newline > Questions what decisions dashboard should support, whether current \org{metrics} align with \pnk{business questions}, and principles of alert/exception highlighting 
& \textit{> Violin plot design critique} \newline > Provides specific enhancements: \cyn{KPI cards with trend sparklines, color-coded performance indicators (red/yellow/green), clearer time period selectors, and drill-down capabilities}. \\
\specialrule{0.8pt}{2pt}{2pt}

P14 & \multicolumn{2}{p{13cm}}{\textbf{Context:} Regional performance charts \newline \textbf{Question:} \textit{``How can I redesign the charts so that each region's performance is clearer and easier to compare?''}} \\
\cmidrule(lr){2-3}
& \textit{> Temporal scale and event overlap principles} \newline > Focuses on \org{temporal scale principles}: whether \org{linear vs. logarithmic time better} serves the story, and how to handle overlapping events 
& \textit{> Swim lanes and visual improvements} \newline > Suggests specific improvements: \cyn{swim lanes for parallel events, color coding by event type.} \\
\specialrule{0.8pt}{2pt}{2pt}

P15 & \multicolumn{2}{p{13cm}}{\textbf{Context:} Agreement percentage bar chart with Kappa values \newline \textbf{Question:} \textit{``The overall message isn't very clear - how can I make patterns more immediately obvious?''}} \\
\cmidrule(lr){2-3}
& \textit{> Radar vs. alternatives comparison principles} \newline > Questions which dimensions matter most for \org{comparison}, whether \cyn{current chart type} (radar) facilitates \org{effective comparison}, and \cyn{principles of parallel coordinates vs. small multiples} 
& \textit{> Grouping and annotation recommendations} \newline > Recommends specific alternatives: \cyn{grouped bar charts for easier comparison and consistent color usage} \\
\specialrule{0.8pt}{2pt}{2pt}

P16 & \multicolumn{2}{p{13cm}}{\textbf{Context:} WordStream authoring pipeline \newline \textbf{Question:} \textit{``Can you give me recommendations on increasing ease of use for this no-code tool?''}} \\
\cmidrule(lr){2-3}
& \textit{> Workflow smoothness principles} \newline > Focuses on \org{geographic aggregation principles}: whether administrative boundaries align with phenomena, and \cyn{choropleth vs. proportional symbol encoding}
& \textit{> Setup panel usability improvements} \newline > Suggests specific map improvements: better \cyn{color scale (sequential vs. diverging), legend with clear thresholds, adding regional labels, and supplementary bar chart} for ranked comparison \\
\specialrule{0.8pt}{2pt}{2pt}

P17 & \multicolumn{2}{p{13cm}}{\textbf{Context:} Sales and Profit side-by-side bar charts \newline \textbf{Question:} \textit{``Is there a more compact way to plot these for easier comparison without losing the message?''}} \\
\cmidrule(lr){2-3}
& \textit{> Chart comparison principles} \newline > Questions what textual patterns matter for \pnk{research question}, whether word clouds serve \pnk{analytical goals}, and principles of \org{text summarization vs. raw frequency}. 
& \textit{> Specific compact alternatives} \newline > Provides specific alternatives to word cloud: \cyn{horizontal bar chart of top terms, topic evolution over time line chart, interactive filtering by topic, and example quotes for context} \\
\specialrule{0.8pt}{2pt}{2pt}

P18 & \multicolumn{2}{p{13cm}}{\textbf{Context:} Parking lot price ranking table \newline \textbf{Question:} \textit{``Do you have recommendations for improving the order of parking lot prices for ranking?''}} \\
\cmidrule(lr){2-3}
& \textit{> Dimensionality reduction principles} \newline > Focuses on \pur{dimensionality reduction} principles: clustering to reveal structure, whether full matrix serves \pnk{exploration goal}, and \pur{threshold-based filtering} concepts 
& \textit{> Clustering and ordering suggestions} \newline > Suggests specific improvements: clearer \cyn{color scale with diverging colors, and interactive hover for exact values} \\
\specialrule{0.8pt}{2pt}{2pt}

P19 & \multicolumn{2}{p{13cm}}{\textbf{Context:} Wide bar chart with dynamic zone visibility \newline \textbf{Question:} \textit{``The bar chart is intentionally wide - what are better ways to manage clickable city details?''}} \\
\cmidrule(lr){2-3}
& \textit{> Aspect ratio and pattern principles} \newline > Questions what patterns comparison should reveal \org{(trends, anomalies, relative positions)}, and \cyn{principles of aspect ratio for time series perception}
& \textit{> Small multiples and layout solutions} \newline > Recommends specific solutions: \cyn{small multiples with shared scales, highlighting reference series, adding annotations for key events, and range selectors for zooming} \\
\specialrule{0.8pt}{2pt}{2pt}

P20 & \multicolumn{2}{p{13cm}}{\textbf{Context:} Thyroid hormone tracking over time \newline \textbf{Question:} \textit{``What visualization approach would help viewers quickly see TSH/T3/T4 trends and treatment changes?''}} \\
\cmidrule(lr){2-3}
& \textit{> Distribution vs. summary statistics principles} \newline > Focuses on whether showing distributions vs. summary statistics better serves the comparison \pnk{goal}, and \cyn{principles of diverging stacked bars vs. separate positive/negative} 
& \textit{> Diverging visualization recommendations} \newline > Suggests specific chart type: \cyn{diverging stacked bar with neutral in center, sorting by agreement level, clear question labels, and percentage labels for each segment} \\
\specialrule{0.8pt}{2pt}{2pt}

P21 & \multicolumn{2}{p{13cm}}{\textbf{Context:} Complex visualization (unspecified) \newline \textbf{Question:} \textit{``Is this visualization difficult to read? What should I change to make it easier?''}} \\
\cmidrule(lr){2-3}
& \textit{> Abstraction hierarchy principles} \newline > Questions what decisions flowchart should support, whether all detail level is necessary, and \org{principles of abstraction hierarchies}
& \textit{> Simplification strategies} \newline > Provides specific simplification strategies: \org{grouping steps into phases}, \cyn{clearer decision diamonds, color coding by department/role}, and \org{collapsible detail sections} \\
\specialrule{0.8pt}{2pt}{2pt}

P22 & \multicolumn{2}{p{13cm}}{\textbf{Context:} Group trends comparison (pre vs. during pandemic) \newline \textbf{Question:} \textit{``How can I more clearly show how trends differ across groups before vs. during the pandemic?''}} \\
\cmidrule(lr){2-3}
& \textit{> Hierarchical comparison principles} \newline > Focuses on \org{hierarchical vs. flat comparison principles}, whether area encoding serves the \pnk{goal}
& \textit{> Same principles discussion} \newline > Focuses on when \cyn{treemap vs. sunburst vs. bar chart} is appropriate \\
\specialrule{0.8pt}{2pt}{2pt}

P23 & \multicolumn{2}{p{13cm}}{\textbf{Context:} Group trends comparison (pre vs. during pandemic) \newline \textbf{Question:} \textit{``How can I more clearly show how trends differ across groups before vs. during the pandemic?''}} \\
\cmidrule(lr){2-3}
& \textit{> Network property principles} \newline > Questions what \pur{network properties matter for analysis (centrality, communities, paths)}, and \cyn{principles of node-link vs. matrix vs. arc diagrams} 
& \textit{> Node sizing and filtering recommendations} \newline > Recommends specific enhancements: \org{size nodes by connection count, color by community detection, label only important nodes.} \\
\specialrule{0.8pt}{2pt}{2pt}

P24 & \multicolumn{2}{p{13cm}}{\textbf{Context:} Semantic prompt prefix diverging bar chart \newline \textbf{Question:} \textit{``How can readers immediately grasp the semantic differences between P1 and P2?''}} \\
\cmidrule(lr){2-3}
& \textit{> Information architecture principles} \newline > Focuses on \org{information architecture principles: logical grouping by decision workflow, visual hierarchy for importance}, and \cyn{principles of dashboard layout (F-pattern, Z-pattern)} 
& \textit{> Layout redesign with visual hierarchy} \newline > Provides specific layout redesign: \cyn{clear sections with headers, primary metrics in top cards, related charts grouped together, consistent chart styles, and better spacing/whitespace} \\

\bottomrule
\end{longtable}

\end{document}